\documentclass[twocolumn]{article}

\pdfoutput=1 

\usepackage{cite}
\usepackage[colorlinks=true, linkcolor=blue, citecolor=blue, urlcolor=blue]{hyperref}
\usepackage{amsmath}
\usepackage{amssymb}
\usepackage{float}
 \usepackage{fullpage}
 \usepackage{rotating}
 \usepackage{multicol}
 \usepackage{array}
\usepackage[table]{xcolor}
 \usepackage[abs]{overpic}
 \usepackage{mathptm,graphicx,rotate,color}
 \definecolor{lightgray}{gray}{0.5}

\begin{document}

\noindent{\Large{\textbf{Happiness and the Patterns of Life: A Study of Geolocated Tweets}}
\normalsize

\noindent{Morgan R. Frank, Lewis Mitchell, Peter Sheridan Dodds$^{\#}$, \\ Christopher M. Danforth$^{*}$ \\ Scientific Reports. 2013, \textbf{3}, No: 2625, doi:10.1038/srep02625

\footnotesize

\noindent{Computational Story Lab, Department of Mathematics and Statistics, Vermont Complex Systems Center, Vermont Advanced Computing Core, University of Vermont, Burlington, Vermont, United States of America
$^{\#}$peter.dodds@uvm.edu, $^{*}$chris.danforth@uvm.edu (corresponding authors) \\

\normalsize

\noindent{{\bf The patterns of life exhibited by large populations have been described and modeled both as a basic science exercise and for a range of applied goals such as reducing automotive congestion, improving disaster response, and even predicting the location of individuals. However, these studies have had limited access to conversation content, rendering changes in expression as a function of movement invisible. In addition, they typically use the communication between a mobile phone and its nearest antenna tower to infer position, limiting the spatial resolution of the data to the geographical region serviced by each cellphone tower. We use a collection of 37 million geolocated tweets to characterize the movement patterns of 180,000 individuals, taking advantage of several orders of magnitude of increased spatial accuracy relative to previous work. Employing the recently developed sentiment analysis instrument known as the \textit{hedonometer}, we characterize changes in word usage as a function of movement, and find that expressed happiness increases logarithmically with distance from an individual's average location.}
%

\indent A proper characterization of human mobility patterns \cite{semini,movement,song,nathaneagle,Montjoye,barabasi,scaling,bagrow2012,levy2,group,dynamics,burst,habitual,press,brownian,agents} is an essential component in the development of models of urban planning \cite{urban}, traffic forecasting \cite{traffic}, and the spread of diseases \cite{diseaseSpread,epidemics,forecast}. In the modern communication era, patterns of human movement have been revealed at an increasingly higher resolution in both space and time, with mobile phone data in particular complementing existing survey-based investigations. As is the case with each new instrument measuring macroscale sociotechnical phenomena, the task has become one of understanding what discernible patterns exist, and what meaning can be derived from those patterns \cite{nathaneagle,election,flashCrash,predict}. \\
\indent Scientists working to understand mobility have employed a diverse set of methodologies. Brockmann et al. \cite{scaling} used the circulation of nearly 1/2 million U.S. dollar bills whose locations were submitted by over 1 million visitors to a website \cite{wheresgeorge} to demonstrate that bank note trajectories are superdiffusive in space and subdiffusive in time, i.e. moving farther and less frequently than expected. \\
\indent Gonzalez et al. \cite{movement} used 6 months of mobile phone data from 100,000 individuals to show that human trajectories are regular in space and time, with each individual having a high probability of returning to a few preferred locations according to Zipf's law. Combining phone communication data with measures of community economic prosperity, Eagle et al. \cite{nathaneagle} showed that the diversity of contacts in an individual's social network is strongly correlated to the potential for economic development exhibited by their community. Finally, de Montjoye et al. \cite{Montjoye} recently used mobile phone data to show that four space-time locations are enough to uniquely identify 95\% of individuals.  \\
\indent Exemplifying recent work to characterize sentiment with social network communications, Mitchell et al. \cite{Lewis} combined traditional survey data (e.g., Gallup) with millions of tweets to correlate word usage with the demographic characteristics of U.S. urban areas. Expressed happiness was shown, for example, to correlate strongly with percentage of the population married, and anti-correlate with obesity. Words such as ``McDonald's" and ``hungry" appeared far more frequently in obese cities, suggesting their instrument could be used to provide real-time feedback on social health programs such as the proposed ban on the sale of large sodas in New York City in 2013.\\
\indent In what follows, we characterize the pattern of life of over 180,000 individuals mainly in the U.S. using messages sent via the social networking service Twitter, and employ our text-based \textit{hedonometer} \cite{hedono} to characterize sentiment as a function of movement. In the calendar year 2011, we collected roughly 4 billion messages through Twitter's gardenhose feed, representing a random 10\% of all status updates posted during this period. \\
\indent Along with an abundance of other metadata, location information typically accompanies each message, resulting from one of three mechanisms by which individuals can report their location when updating their status. First, when an individual registers their account with Twitter, they are presented with the opportunity to report their location in a free text box. This region will be displayed in their user profile (e.g. `NYC' or `over the rainbow'). The metadata accompanying each tweet sent by the individual contains this self-reported location. Second, individuals submitting a message through a web browser can choose to tag their message with a `place' chosen from a drop-down menu, where the first option provided is typically the city within which the computer's IP address is found. For the purposes of accuracy, we have chosen to ignore each of these two mechanisms for reporting position when attempting to assign each tweet a geographical location, and focus instead on messages located via a third mechanism, namely the Global Positioning System (GPS). \\
\indent Individuals using a mobile device application may opt-in to \textit{geolocate} their message, in which case the exact latitude and longitude of the mobile phone is reported. The accuracy of this information is governed by the precision of the GPS instrument embedded in the phone, which can vary depending on the surrounding topography. As a result of these factors, we are able to approximately place each geolocated message inside a 10 meter circle on the surface of the Earth, within which the tweet was sent. Roughly $1\%$ of the status updates received through the gardenhose feed are geolocated, resulting in a total of 37 million messages, collectively representing more than 180,000 English-speaking people worldwide. Fig. \ref{Figdots} illustrates the geospatial resolution of the data. \\
\begin{sidewaysfigure*}[!htdp]
\begin{center}
	\includegraphics[width=\textwidth]{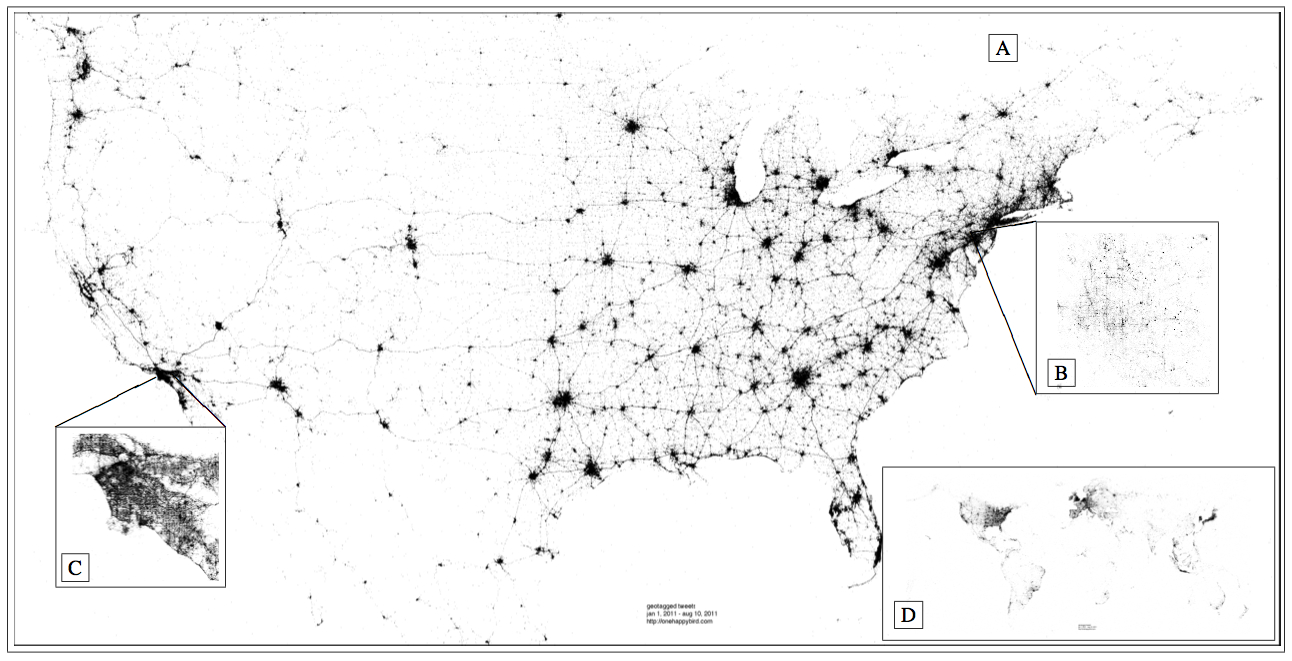}
%
	\end{center}
	\caption{Each point corresponds to a geolocated tweet posted in 2011. Twitter activity is most apparent in urban areas. Note that the image contains no cartographic borders, simply a small dot for each message.  Legend: A (U.S.), B (Washington, D.C.), C (Los Angeles, C.A.),  and D (Earth).}
\label{Figdots} 
\end{sidewaysfigure*}
\noindent{\textbf{Results}} \\
\indent Following Gonz\'{a}lez et al.\cite{movement}, we examine the shape of human mobility using \emph{radius of gyration}, hereafter gyradius, as a measure of the linear size occupied by an individual's trajectory. In Fig. \ref{Figradcitydots}, we investigate the geographical distribution of movement in four urban areas by plotting a dot for each tweet, colored by the gyradius of its author. Clockwise from the top left, cities are displayed in order of their apparent aggregate gyradius, with New York City seemingly exhibiting a smaller radius than the San Francisco Bay Area. In Chicago, many individuals writing from downtown exhibit an order of magnitude greater radius than individuals posting in areas outside of the city. A similar pattern is seen when looking at each point colored instead by distance from expected location (Fig. \ref{Figcitydots2})).\\
\begin{figure*}[!htdp]
\begin{center}
	\includegraphics[width=\textwidth]{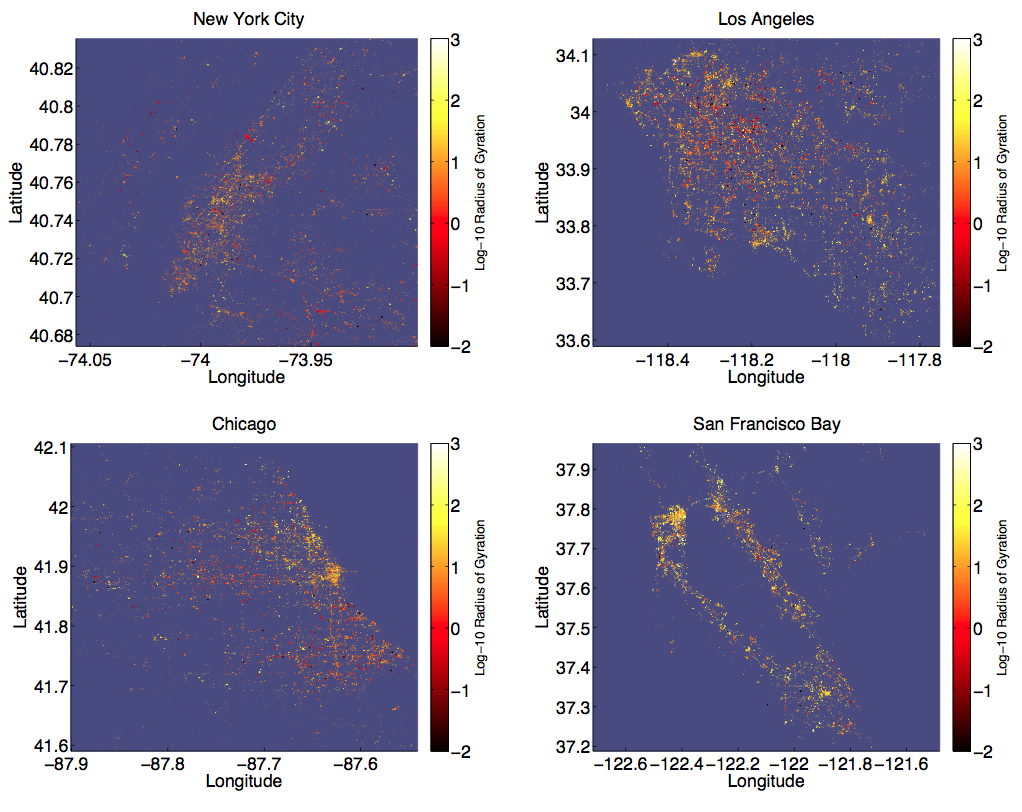}
\end{center}
\caption{The gyradius, calculated for each individual, is shown for each tweet authored in four example cities. Tweet activity reflects population density, with urban areas clearly visible in each city. Histograms of gyradii for each city are shown in Fig. \ref{Figraddist}, along with tweet locations colored by distance from expected location (Fig. \ref{Figcitydots2}). The number of tweets shown for each city is $N=42,089$ (New York City), $N=103,213$ (Los Angeles), $N=56,650$ (Chicago), and $N=45,754$ (San Francisco). Note that higher resolution versions of the four panels above can be found online \cite{frankonline}.}
\label{Figradcitydots} 
\end{figure*}
\indent In the greater Los Angeles area, we see several clusters of individuals with larger radius in downtown Los Angeles, as well as Long Beach, Santa Monica, and Disneyland in Anaheim, while less densely populated areas are seen as smaller clusters exhibiting much smaller radii. The geography of the San Francisco Bay Area is clearly revealed, with many large radius individuals tweeting from downtown San Francisco, and somewhat less homogeneity in Oakland and San Jose. Outside of these cities, there are many suburban areas revealed by individuals with large radius, e.g. Palo Alto. Tweets appearing in less densely populated Bay Area locations appear to be far more likely to be authored by large radius individuals than those appearing in lower population areas of the other cities. This observation likely reflects the socio-economic and demographic characteristics of individuals using Twitter in the Bay Area, where the social network service was founded. Additionally, it could reflect the presence of tourists who will typically have a larger radius than someone who lives and works in the Bay Area.

\begin{figure*}[!htdp]
\begin{center}
\includegraphics[width=.85\textwidth,trim=3cm 0cm 3cm 0cm,clip]{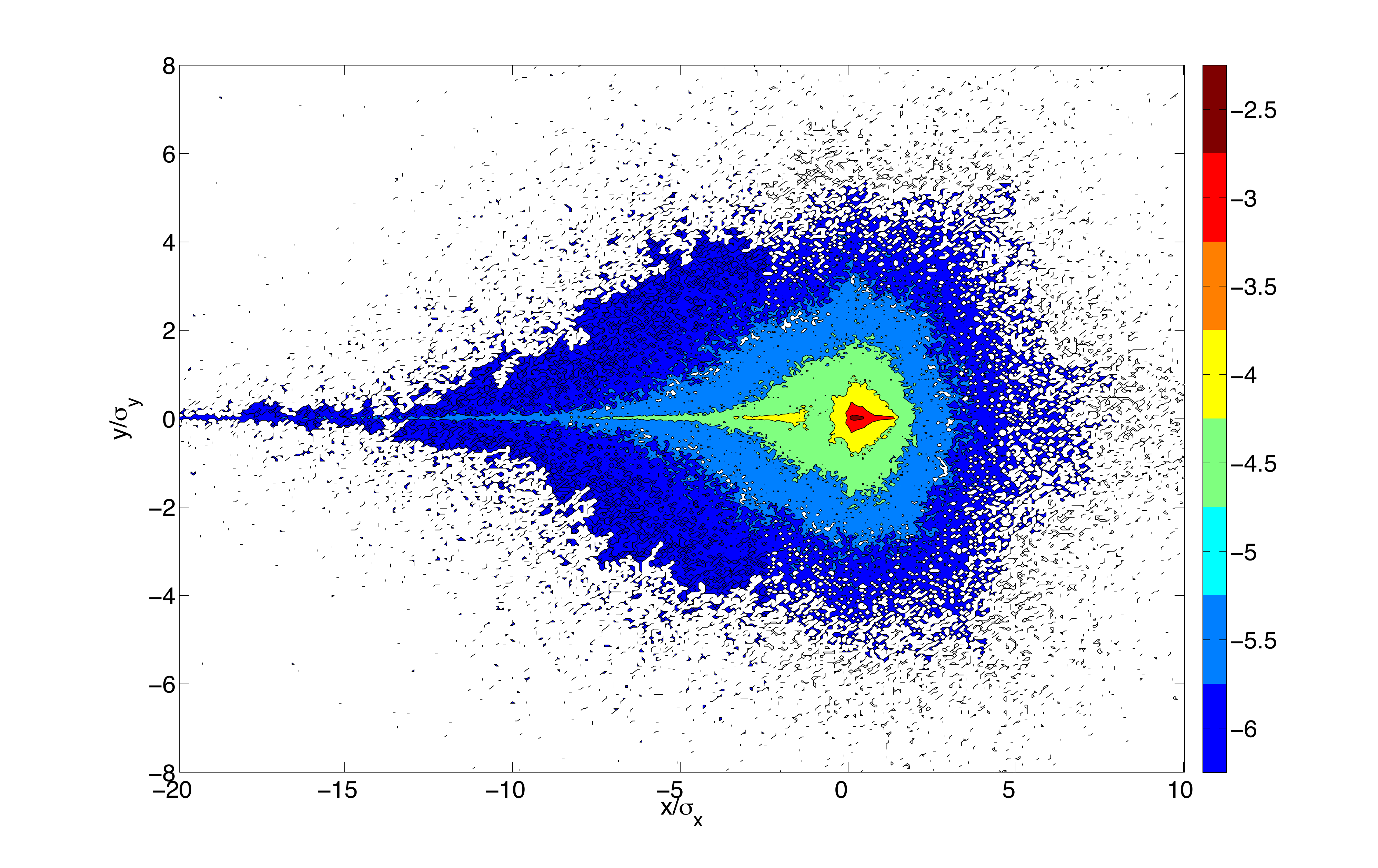}
		\end{center}
\caption{The probability density function of observing an individual in their normalized reference frame, where the origin corresponds to each individual's expected location, and $\sigma_{y}$ = 0 corresponds to their principle axis. This map shows the positions of over 37,000 individuals, each with more than 50 locations, in their intrinsic reference frame.}
\label{Figdensity} 
\end{figure*}
\indent We calculate Geary's C (local) and Moran's I (global) spatial autocorrelation for the data shown in Figures \ref{Figradcitydots} and \ref{Figcitydots2}, finding statistical support for spatial clustering in each (Tables \ref{Taspatial1} and \ref{Taspatial2}). However, the correlations benefit from the propensity for each individual's collection of tweets to exhibit clustering. To avoid this confound, we also make city plots of mode location colored by gyradius, where each dot represents an individual rather than a tweet. These figures are not included to respect the privacy of individuals in the study. Table \ref{Taspatial3} reports the strong spatial autocorrelation we observed, reflecting a form of geospatial homophily: the tendency of individuals to author messages in proximity to others with similar gyradius. Tourists are unlikely to be included in this statistic, given the nature of mode location, and as such the clustering is potentially a result of similar commute distances.\\
\indent One observation seemingly apparent in Fig. \ref{Figradcitydots} is that individuals who move a lot tend to appear in areas of large population density. Given the apparent economies of scale offered by living in a densely populated area, one might expect to observe the inverse relationship, namely that people living in less densely populated areas travel further, by necessity, to their place of employment or grocery store, for example. Of course, individuals observed to have a large radius could be tourists, or they could have a long commute. Nevertheless, we find no statistical evidence for this trend. Comparing individuals whose average location falls in an area of small vs. large tweet density, we observe little difference in their average gyradii (not shown). \\
\indent Moving beyond these four urban areas and looking at 472 cities in the U.S., we do find a moderate correlation between the mean gyradius and city land area (Pearson $\rho=0.24, \ p=2\times10^{-7}$); Fig. \ref{Figpopulation} and Table \ref{Tacities} show the top and bottom cities with respect to gyradii.\\
\indent To investigate the shape of human mobility, we normalize each individual's trajectory to a common reference frame (see Methods). In Fig. \ref{Figdensity}, we plot a heat map of the probability density function of the normalized locations of all individuals. For the purposes of this discussion, we will refer to deviations from an individual's expected location in the normalized reference frame as occurring in the directions north, south, east, and west. Several features of the map reveal interesting patterns of movement. First, the overall west-to-east teardrop shape of the contours demonstrates that people travel predominantly along their principle axis, namely heading west from the origin along $y/\sigma_{y}$ = 0, with deviations in the orthogonal direction becoming shorter and less frequent as they move farther away from the origin.  

\begin{figure*}[!htdp]
	\includegraphics[width=\textwidth]{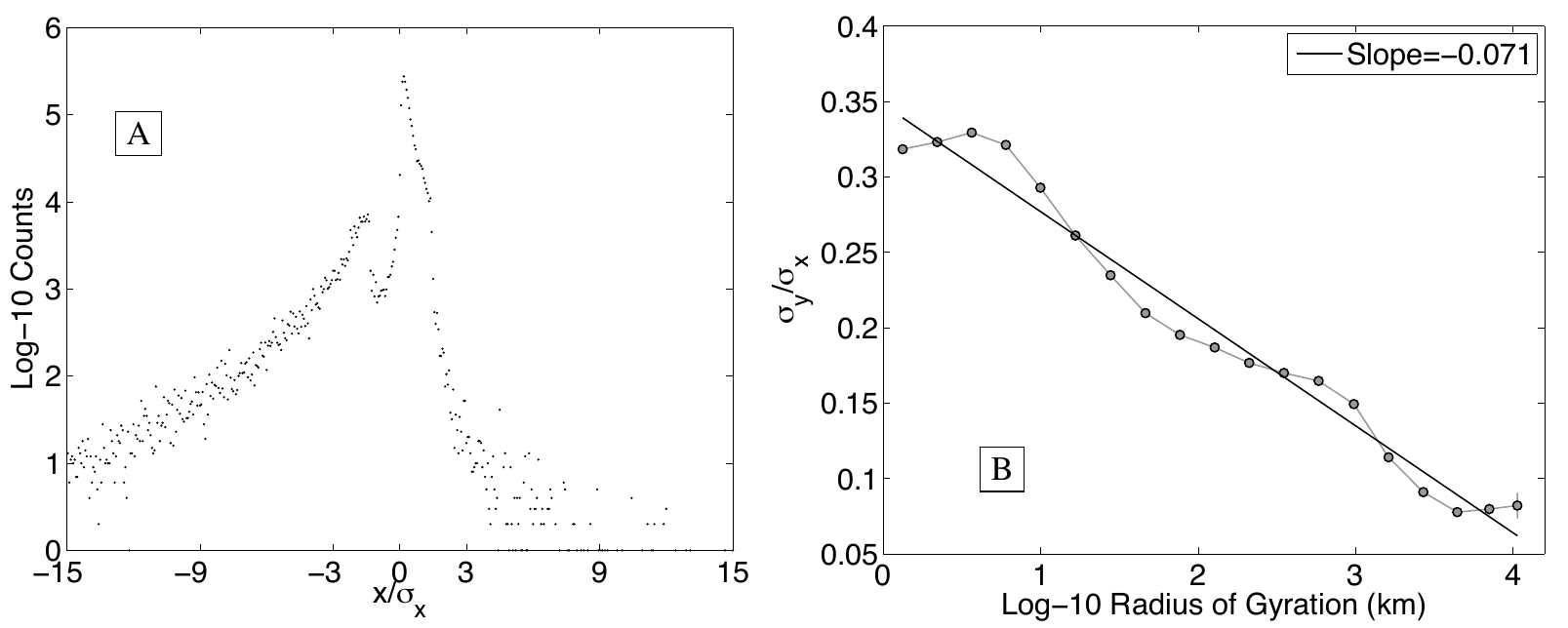}
\caption{Looking at messages authored in the principle axis corridor, defined by $|\frac{y}{\sigma_y}|<\frac{30}{1000}$, we observe a clear separation between the most likely and second most likely position (A). The distribution is skewed left, with movement in a heading opposite an individual's work/home corridor observed to be highly unlikely. In addition, due to the normalization, we see that individuals are much more likely to tweet slightly east of their expected location than slightly west. The isotropy ratio (B) measures the change in the density's shape as a function of gyradius, with large radius individuals exhibiting a less circular pattern of life. Standard errors are plotted, but are only visible for the largest radius group. The isotropy ratio decays logarithmically with radius.}
\label{Figisotropy} 
\end{figure*}
\indent Second, the appearance of two spatially distinct yellow regions separated by a less populated green region suggests that people spend the vast majority of their time near two locations. We refer to these locations as the \textit{work} and \textit{home} locales \cite{bagrow2012}, where the home locale is centered on the dark red region roughly 1 standard deviation east of the origin, and the work locale is centered approximately 2 standard deviations west of the origin. These locations highlight the bimodal distribution of principal axis corridor messages (Fig. \ref{Figisotropy}A). \\
\indent Finally, a clear asymmetry is observed about the $x/\sigma_{x} = 0$ axis indicating the increasingly isotropic variation in movement surrounding the home locale, as compared to the work locale. We interpret this to be a reflection of the tendency to be more familiar with the surroundings of one's home, and to explore these surroundings in a more social context (Fig. \ref{Figisotropy}B). The symmetry observed when reflecting about the $y/\sigma_{y} = 0$-axis is strong, demonstrating the remarkable consistency of the movement patterns revealed by the data.\\
\indent In an effort to characterize the temporal and spatial structure observed in Fig. \ref{Figdensity}, in Fig. \ref{Figzipf} we examine locations frequently visited by the most active members of our data set, namely the roughly 300 individuals for whom we received at least 800 geolocated messages. We suspect that these individuals enabled the geolocating feature to be \textit{on} by default for all messages, as implied by the roughly $O(10^4)$ geolocated messages suggested by the gardenhose rate. In Fig. \ref{Figzipf}, we focus on these individuals specifically; of all participants, their prolific tweet activity most accurately reflects their movement profile.\\
\indent The main figure shows the probability of tweeting from each locale, with locales ordered by rank, for each individual \cite{bagrow2012}. We find that $P(H_{i}^{(a)})\propto R(H_{i}^{(a)})^{-1.3}$ which is approximately a Zipf distribution \cite{zipf}. This finding indicates that regardless of the number of tweet locales for a given individual, the majority of their messaging activity occurs in one of only a few locales, with the probability decaying at a predictable rate. If the decay were Zipfian, an individual would be approximately $n$-times as likely to tweet from their mode location than from their rank $n$ location. With our slope being steeper, these probabilities fall at a faster rate with rank. The slope is robust to variation in number and composition of individuals (Fig. \ref{Figzipfstats}).\\
\indent For roughly 95\% of these individuals, each tweet has a greater than 10\% chance of being authored from their mode location (Fig. \ref{Figzipf}B). Fig. \ref{Figzipf}C demonstrates each individual's likelihood of authoring messages from their mode location (black curve) at different times of day throughout the week. A period-2 cycle is observed for each day of the week. Maxima are seen in the morning (8-10am) and evening (10pm-midnight), and minima in the afternoon (2-4pm) and overnight (2-4am) hours. The peak in the morning is consistently higher than that in the evening, and the afternoon valley is consistently lower than the overnight valley. The cycle is somewhat less structured on the weekend. Also plotted are the probabilities of tweeting from locations other than the mode (red curve).\\
\indent In a study performed with cellphone tower data, Gonz\'{a}lez et al. \cite{movement} found that people spend most of their time in two locations, and a person's probability of being found at a separate location diminishes rapidly with rank by visitation. While our investigation reveals a similar pattern, we find a larger difference in the probability that an individual is tweeting from the home locale than from the work locale. We attribute these slight differences in our results to the different spatiotemporal precision of location data, as well as differences in activities represented by the data. Gonz\'{a}lez et al. determined each individual's location by continuously monitoring the nearest cellphone tower whose range they were within. As such, we receive more precise location information, but only when individuals performed the act of tweeting.\\
\begin{figure*}[!htdp]
\begin{center}
	\includegraphics[width=\textwidth]{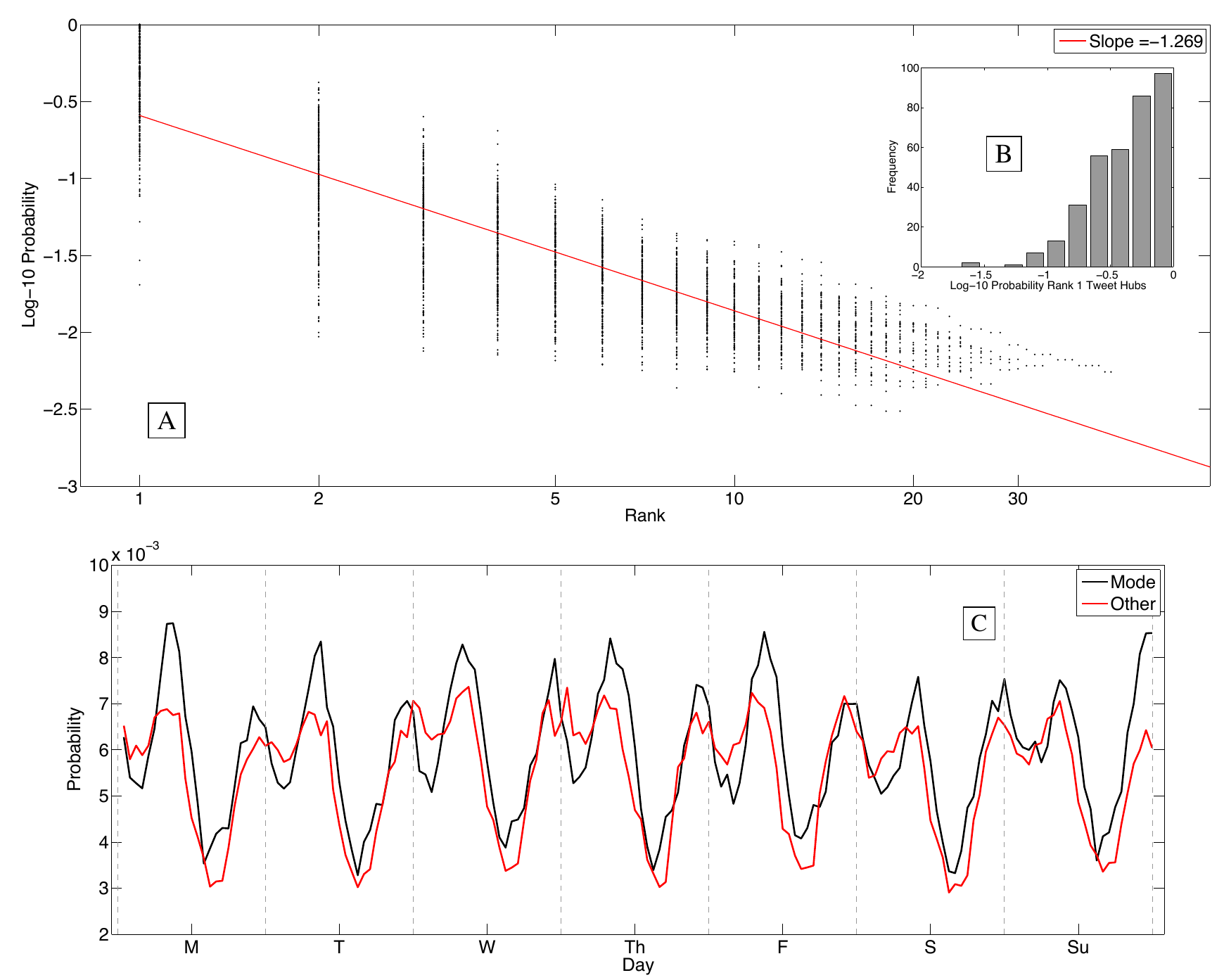}
\end{center}
\caption{Representing the approximately 300 individuals for whom we have at least 800 geolocated messages, we plot the probability of tweeting from a locale as a function of the tweet locale rank (A). Each dot represents a single individual's likelihood of tweeting from one of their locales. The axes are logarithmic, revealing an approximate Zipfian distribution with slope -1.3 \cite{zipf}. (B) Distribution of the rank-1 locale, each individual's mode location. (C) A robust diurnal cycle is observed in the hourly time of day at which statuses are updated, with those from the mode location (black curve) occurring more often than other locations (red curve) in the morning and evening. Probabilities sum to 1 for each curve, with bins for each hour. Dashed vertical lines denote midnight.}
\label{Figzipf} 
\end{figure*}
\begin{figure*}[!htdp]
\begin{center}
	\includegraphics[width=\textwidth]{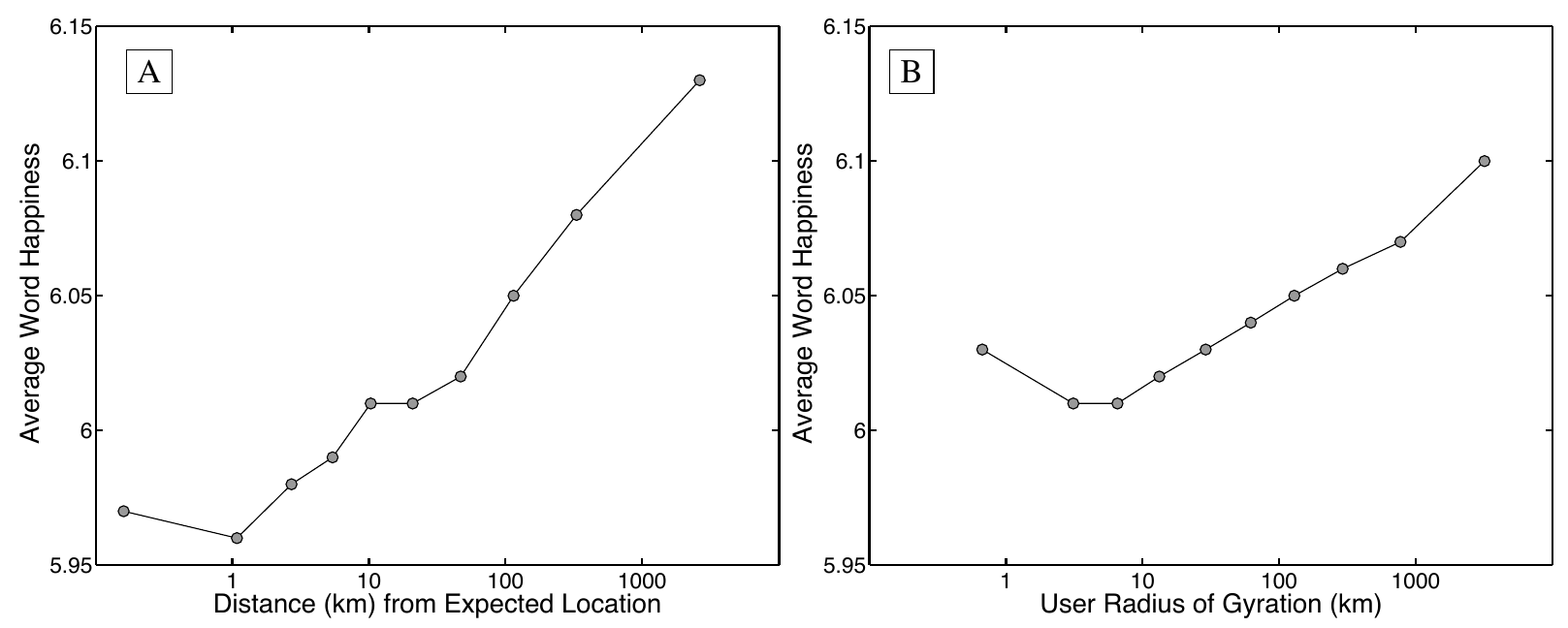}
\end{center}
\caption{(A) Average happiness of words written as a function of distance from an author's expected location, with tweets grouped into ten equally populated bins. Expressed happiness grows logarithmically with distance distance from expected location. (B) A similar trend is observed when individuals are grouped into ten equally populated bins according to their gyradius. Both trends persist through variations in binning and different measures of mobility.}
\label{Fighappy} 
\end{figure*}
\indent One major advantage of using Twitter data to study movement is the additional source of information provided by the messages themselves. Researchers using mobile phone data to characterize mobility patterns do not have access to conversations occurring during the time period of interest. To measure the sentiment associated with different patterns of movement, we use the hedonometer introduced by Dodds et al. \cite{hedono}. The instrument performs a context-free measurement of the happiness of a large collection of words using the language assessment by Mechanical Turk (labMT) word list, as described in Kloumann et al. \cite{positivity}. LabMT comprises roughly 10,000 of the most frequently used words in the English language, each of which was scored for happiness on a scale of 1 (sad) to 9 (happy) by people using Amazon's Mechanical Turk service \cite{amazon,mecturk}, resulting in an average happiness score for each word. Example word scores are shown in Table \ref{labMTexamples}. Note that in employing the hedonometer, we avoid assigning sentiment to individual tweets, a challenging task more appropriately suited to advanced natural language processing software. \\
\indent To examine the relationship between movement and happiness, we calculate expressed happiness as a function of distance from an individual's expected location, as well as gyradius. For the former, we grouped tweets into ten equally populated bins, with each group containing more than 500,000 tweets from similar distances. The happiness of each group was then computed using Eqn \ref{Eq:labMT} (see Methods), where all words written from a given distance were gathered into a single bin. For the latter, we placed individuals into ten equally sized groups by gyradius, with each group containing more than 10,000 individuals with similar gyradii. \\
\indent Fig. \ref{Fighappy} plots average word happiness against the distance from expected location (A), and gyradius (B). Starting with location, we find that tweets written close to an individual's center of mass are slightly happier than those written 1km away. The least happy words, on average, are used at a distance representative of a short daily commute to work. Beyond this least happy distance, remarkably we find that happiness increases logarithmically with distance from expected location. Perhaps even more remarkably, we find an almost identical trend when grouping together individuals rather than tweets, observing that happiness also increases logarithmically with gyradius. Individuals with a large radius use happier words than those with a smaller pattern of life. We find the trend observed in Fig. \ref{Fighappy} holds for 3 of the 4 urban areas (Los Angeles, San Francisco, and Chicago), see Figs. \ref{Fighappy1}, \ref{Fighappy2}. \\
\indent To explain the difference in expressed happiness exhibited by different mobility groups, we turn to word shift graphs in Fig. \ref{Fighappywords}. Word shift graphs were introduced by Dodds and Danforth \cite{largeScale,hedono} as a means for investigating the elements of language responsible for happiness differences between two large texts. As an example, consider the difference between tweets authored at distances of roughly 1km and 2500km away from an individual's expected location. The average happiness scores for these two distances are $h_{avg} = 5.96$ and $h_{avg} = 6.13$ respectively. Individual word contributions to this difference are shown in Fig. \ref{Fighappywords}A, and can be described as follows.\\
\indent Words appearing on the right increase the happiness of the 2500km distance relative 1km distance. For example, tweets authored far from an individual's expected location are more likely to contain the positive words `beach', `new', `great', `park', `restaurant', `dinner', `resort', `coffee', `lunch', `cafe', and `food', and less likely to contain the negative words `no', `don't', `not', `hate', `can't', `damn', and `never' than tweets posted close to home. Words going against the trend appear on the left, decreasing the happiness of the 2500km distance group relative to the 1km group. Tweets close to home are more likely to contain the positive words `me', `lol', `love', `like', `haha', `my', `you', and `good'. Moving clockwise, the three insets in Fig. \ref{Fighappywords}A show that the two text sizes are comparable, the biggest contributor to the happiness difference is the decrease in negative words authored by individuals very far from their expected location, and the 50 words listed make up roughly 50\% of the total difference between the two bags of words. \\
\indent Note that the relatively small differences in $h_{avg}$ scores reflect a small signal, yet one that we have shown previously can be resolved by our hedonometer \cite{hedono}. Additional word shift comparisons for the four urban areas investigated earlier are provided in the Supplemental Material, Figs. \ref{Figwords1}, \ref{Figwords2}. \\
\indent Looking at the word differences between individuals with largest and smallest radii of gyration in Fig. \ref{Fighappywords}B, we see that individuals in the large radius group author the negative words `hate', `damn', `dont', `mad', `never', `not' and assorted profanity less frequently, and the positive words `great', `new', `dinner', `hahaha', and `lunch' more frequently than the small radius group. Going against the trend, the large radius group uses the positive words `me', `lol', `love', `like', `funny', `girl', and `my' less frequently, and the negative words `no', and `last' more frequently. Comparing with other groups, the large radius group authors an increased frequency of words in reference to eating, like the words `dinner', `lunch', `restaurant', and `food', and make less reference to traffic congestion. \\
\indent Comparing the two figures, we note that individuals with large radius laugh more (e.g `hahaha') than those with a small radius, but individuals closer to their expected location laugh more than those far from home. \\
\indent These word differences reveal the relationship between an individual's pattern of movement and their experiences. It is not surprising to observe regular international travelers tweeting about the food they enjoy on vacation. Indeed, we expect that individuals capable of tweeting at a great distance from their expected location are more likely to benefit from an advantaged socioeconomic status, which they happily update frequently. In our earlier work, we have demonstrated that expressed happiness correlates strongly with many socioeconomic indicators \cite{Lewis}. Nevertheless, setting aside these luxurious words, we still see a general decline in the use of negative words as individuals travel farther from their expected location. In fact, of the four contributions to the difference in happiness between words authored close to home vs. far from home, this decline in negative words is the largest component (bottom right inset, Fig. \ref{Fighappywords}). \\
\noindent{\textbf{Discussion}}\\
\indent Using 37 million geolocated tweets authored in 2011, we have been able to characterize the pattern of life of over 180,000 individuals largely residing in the United States. While observed mobility patterns agree qualitatively with previous work investigating cellphone data \cite{movement}, we are able to connect movement patterns to changes in word usage for the first time. Our main finding is that expressed happiness increases logarithmically with both distance from expected location and gyradius, largely because individuals who travel farther use positive, food related words more frequently, and negative words and profanity less frequently. \\ 
\indent Several methodological issues are raised by the use of Twitter messages to characterize mobility and happiness. Considering Twitter as a source, we note that according to the Pew Internet \& American Life Project, roughly 15\% of adults in the U.S. were actively using Twitter at the end of 2011 \cite{pew}. While this fraction represents a substantial group of Americans, we have no data to quantify the demographic group represented by the subset of these 15\% who specifically choose to geolocate a large percentage of their messages. Nevertheless, since we threshold the sample to include individuals who have geolocated more than approximately 300 of their messages in 2011, we suspect that the large majority of individuals represented in our study regularly do so as a matter of daily life, as opposed to geolocating messages only when encountering a novel experience such as a vacation. \\
\indent Regarding word usage as a proxy for happiness, accessing the internal emotional state of individuals is beyond the scope of our instrument. We do believe however, that when aggregated, the words used by large groups of individuals reflect their culture in ways not captured by surveys or self-report. Indeed, we see the hedonometer as complementing more traditional economic methods for characterizing economic and societal health, such as the Gross Domestic Product or Consumer Confidence Index. Using the same collection of geolocated messages explored here, the hedonometer was recently employed by Mitchell et al. \cite{Lewis} to characterize trends in word usage for cities. Expressed happiness was shown to correlate to hundreds of demographic, socio-economic, and health measures, with interactive evidence available in the article's online Appendix \cite{mitchellonline}.\\
\indent Our work contributes to a growing body of literature aimed at observing, describing, modeling, and ultimately explaining the spatiotemporal dynamics of large-scale socio-technical systems. The mobility patterns investigated here could be combined with more traditional surveys (e.g. census data) to inform public policy regarding many important issues, for example relating to the `obesity epidemic' and changes in word usage at the level of individual neighborhoods targeted by public health campaigns. Feedback on society's eating behavior in response to health promotion policies could be available at the level of neighborhoods on a time scale of weeks, in advance of health data outcomes that typically take years. Indeed, epidemiological models of the spread of food-borne illness can now concurrently leverage information about social network connections and geographic proximity \cite{Sadilek}.  \\
\indent In addition, future mental health providers could flag changes in individual behavior revealed through patterns of movement and communication for intervention. For example, a depressed emotional state may be indicated by simultaneously observing marked declines in gyradius, decreased social interactions, and sustained increase in usage of negative words. Natural extensions of this work might combine topological measures of network interactions with geospatial data to predict the likelihood of new links appearing in a social network \cite{onnela}, or to measure the spread of emotions through geographical and topological space \cite{recip}. \\
\noindent{\textbf{Methods}}\\
\indent In an effort at quality control for the geolocated messages, we identified and removed messages posted by robotic accounts and programmed tweeting services designed to automatically send tweets typically not reflecting information about human activity. Preliminary analyses revealed a noticeable presence of bots posting geolocated messages referring to weather, earthquakes, traffic, and coupons. We identified and ignored tweets collected from individuals for whom at least half of their tweets contained any of the words `pressure', `humid', `humidity', `earthquake', `traffic' or `coupon'. \\
\begin{figure*}[!htdp]
	\includegraphics[width=\textwidth]{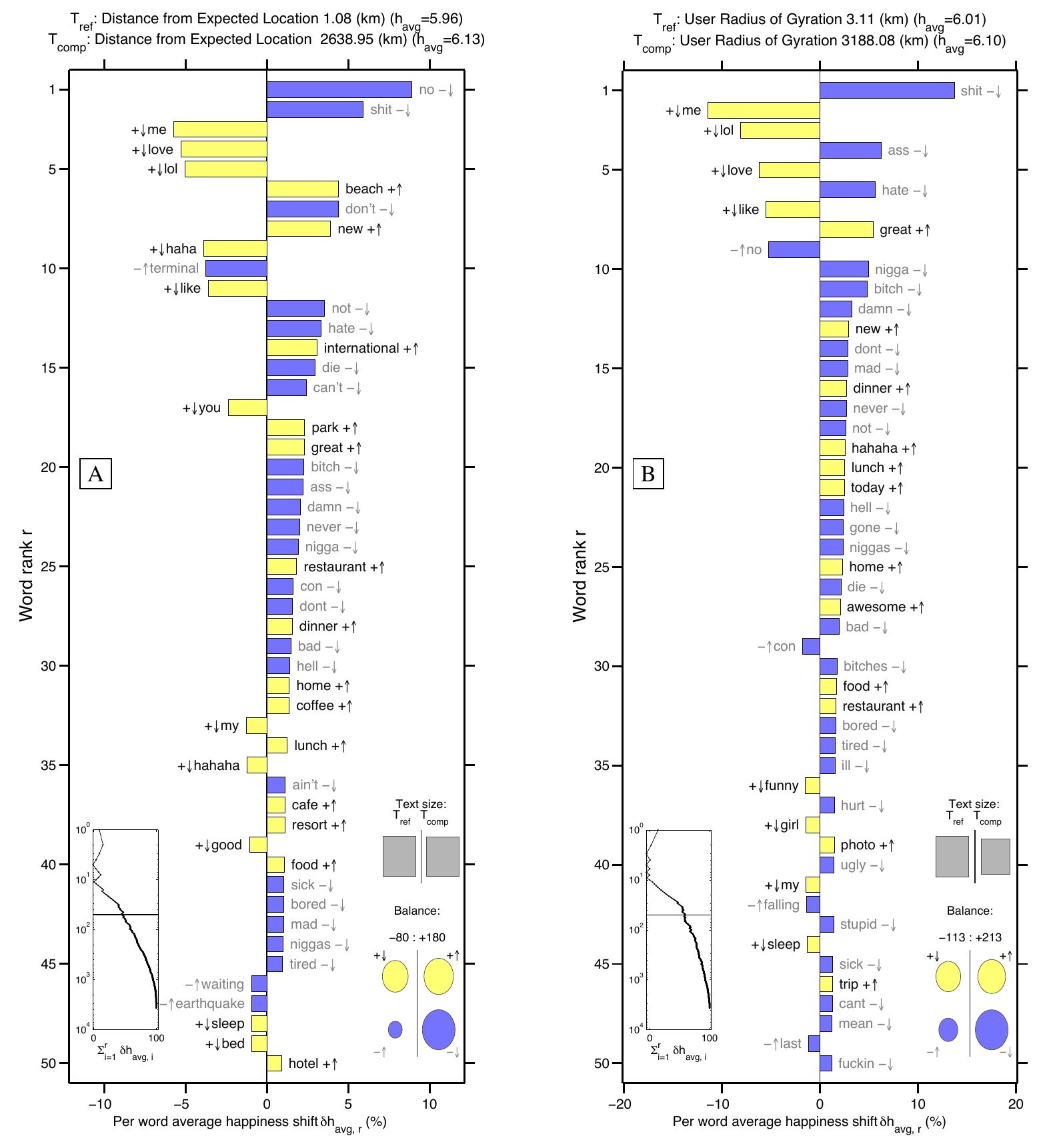}
\caption{Word shift graphs comparing (A) the lowest average word happiness distance from home group to the words authored farthest from home, which also has the largest average word happiness and (B) the smallest gyradius group with the largest gyradius group. The words in the word shifts from top to bottom appear in decreasing order of ranked percentage contribution to the overall average happiness difference ($\Delta h_{avg}$) of the two texts being compared. The +/- symbols indicate whether the word has an average happiness score that is happy or sad relative to the entire text $T_{\rm ref}$. The symbols $\uparrow$ / $\downarrow$ indicate whether a word was used more or less in $T_{\rm comp}$ relative to usage in $T_{\rm ref}$. The left inset panel shows how the ranked top contributing words to $\Delta h_{avg}$ combine in sum. The four circles in the lower right show the total contribution of the four word types $(+\uparrow,-\uparrow,+\downarrow,-\downarrow)$ to the Balance of the happiness difference. The number of words in each of the two texts is represented by the relative area of the grey squares (Text size). See Dodds et al. \cite{hedono} for further details and examples of word shift graphs.}
\label{Fighappywords} 
\end{figure*}
\indent Messages referencing Foursquare check-ins (typically of the form `I'm at starbucks http://4sq.com/qrel9d') were retained for the purpose of characterizing the mobility profile of each individual. However, for results involving happiness, we ignored Foursquare check-in tweets as their content is unlikely to directly reflect sentiment. \\
\indent Finally, to ensure that individual movement profiles are based on a reasonably sized collection of locations, for this study we focus on individuals for whom we have at least 30 geolocated tweets. Given the uniformity of the random sample provided by the gardenhose, we can assume these individuals geolocated a minimum of approximately 300 status updates in 2011. Individuals were included in Figures 6, 7 if their messages matched LabMT words. \\
\indent For reasons of privacy, we ignored all user specific information including individual names. In addition, where the trajectories traced out by specific individuals are visualized, we obscured the coordinate system of reference. Tweets were assigned to urban areas as defined by the 2010 United States Census BureauÕs MAF/TIGER (Master Address File/Topologically Integrated Geographic Encoding and Referencing) database \cite{census}.\\
\indent The gyradius for individual $a$ is defined as
\begin{equation}
	r^{(a)}=\sqrt{\frac{1}{N^{(a)}}\displaystyle\sum_{i=1}^{N^{(a)}}(\vec{p}_{i}^{(a)}-\langle \vec{p}^{(a)}\rangle )^{2}}
\end{equation}\\
where the two-dimensional vector $\vec{p}_{i}^{(a)}$ is the $i$th position in the trajectory of individual $a$, given by the geolocation of that individual's $i$th tweet, as observed in our database. $N^{(a)}$ is the total number of tweets from individual $a$, and $\langle \vec{p}^{(a)} \rangle =1/N^{(a)}\sum_{i=1}^{N^{(a)}}\vec{p}_{i}^{(a)}$ is the center of mass of their trajectory, which we denote their \textit{expected location}. Note that if we consider each message to be a prediction of an individual's location, then the gyradius is in fact the root mean square error (RMSE) of that prediction. Fig. \ref{Figuser5b} plots the Complementary Cumulative Distribution Function (CCDF) of the gyradii of all individuals. \\
\indent To compare the shape of individual trajectories, we normalize for both differences in gyradius and direction of trajectory. Considering each individual's trajectory as a set of $(x,y)$-pairs $\{(x_{1},y_{1}),(x_{2},y_{2}),\dots,(x_{N},y_{N})\}$, we calculate the two dimensional matrix known as the tensor of inertia, considering each point in a individual's trajectory as an equally weighted mass at location $(x_{i},y_{i})$. We then find this tensor's eigenvectors and eigenvalues. The eigenvector corresponding to the largest eigenvalue represents the axis along which most of the individual's trajectory occurs (hereafter called the individual's \emph{principal axis}). Previous work has demonstrated that for most individuals, this axis is parallel to the corridor between their work location and home \cite{movement,song}.\\
\indent To normalize the different compass orientations of individual trajectories, we rotate the coordinate system of each individual so that their principal axis points due west. The expected location for each individual $(\bar{x},\bar{y})$ is then used to translate their position vector, i.e. $(x_{i}-\bar{x},y_{i}-\bar{y})$, to ensure that the shape of each individual's trajectory is in a common frame of reference. However, the distances travelled by each individual vary widely despite their shared orientation (e.g. pedestrian vs. airline commute). In order to compare these trajectories, we calculate the standard deviation $\sigma_{x}$, $\sigma_{y}$ for a given individual's trajectory, and divide their $x$- and $y$-coordinates by $\sigma_{x}$ and $\sigma_{y}$, respectively. For more information about this process, including a pair of example trajectory normalizations, see Figs. \ref{Figuser1}-\ref{Figuser5}.\\
\indent  In an attempt to characterize time spent in each location, we define the $i$th tweet \textit{locale} for individual $a$, denoted $H_{i}^{(a)}$, to be a circle within which individual $a$ posted at least 10 messages \cite{bagrow2012}. The center of the circle is defined by the average position of all messages appearing in the locale, and the radius of the circle is chosen such that each tweet posted within a locale is at most 100 meters away from the center, and no locales overlap.  To measure the importance of locale $i$ to individual $a$, we count the number of messages appearing in each tweet locale and produce the ranking $R(H_{i}^{(a)})$ for individual $a$. The probability that individual $a$ tweets from locale $H_{i}^{(a)}$ is 
\begin{equation}
	P(H_{i}^{(a)})=\frac{|H_{i}^{(a)}|}{N^{(a)}}
\end{equation}
where $|H_{i}^{(a)}|$ is the number of tweet locations contained in $H_{i}^{(a)}$. Notice that the locale probabilities for individual $a$ may not sum to one since it may be the case that individual $a$ has tweet locations that are not contained in a tweet locale. Hereafter, we will refer to an individual's most frequently visited, or rank-1 locale, as their mode location. \\
\indent Using the labMT scores \cite{hedono}, we determine the average happiness ($h_{avg}$) of a given text $T$ containing $N$ unique words by\\
\begin{equation}
	h_{avg}(T)=\frac{\sum_{i=1}^{N}h_{avg}(w_{i})\cdot f_{i}}{\sum_{i=1}^{N}f_{i}}=\sum_{i=1}^{N}h_{avg}(w_{i})\cdot p_{i}
\label{Eq:labMT}
\end{equation}
where $f_{i}$ is the frequency with which the $i$th word $w_{i}$, for which we have an average word happiness score $h_{avg}(w_{i})$, occurred in text $T$. The normalized frequency of $w_{i}$ is then given by $p_{i}=f_{i}/\sum_{i=1}^{N}f_{i}$.\\
\begin{table}[!htdp]
\begin{center}
\begin{tabular}{|c|c|}
\hline
word & $h_{avg}(w_{i})$ \\ \hline
`happy' & 8.30 \\
`hahaha' & 7.94 \\
`fresh' & 7.26 \\
`cherry' & 7.04 \\
`pancake' & 6.96 \\
`piano' & 6.94 \\
\textcolor{lightgray}{`and'} & \textcolor{lightgray}{5.22} \\
\textcolor{lightgray}{`the'} & \textcolor{lightgray}{4.98} \\
\textcolor{lightgray}{`of'} & \textcolor{lightgray}{4.94} \\
`down' & 3.66 \\
`worse' & 2.70 \\
`crash' & 2.60 \\
`:(' & 2.36 \\
`war' & 1.80 \\
`jail' & 1.76 \\ \hline
\end{tabular}
\end{center}
\caption{Example language assessment by Mechanical Turk (labMT) \cite{hedono,positivity} words and scores. Words with neutral scores $4<h_{avg}(w_{i})<6$ are colored gray and ignored when assigning the happiness score to a large text. }
\label{labMTexamples}
\end{table}%
\indent The hedonometer instrument can be tuned to emphasize the most emotionally charged words by removing words within $\Delta h_{avg}$ of the neutral score of $h_{avg} = 5$. We have further shown that ignoring these neutral words with $4<h_{avg}(w_{i})<6$ provides a good balance of sensitivity and robustness, and thus we chose $\Delta h_{avg}=1$ for this study \cite{hedono}.\\

\noindent{\textbf{{Acknowledgements}} \\
\indent For their helpful comments and input, we thank Brian Tivnan, James Bagrow, Yu-Ru Lin, Taylor Ricketts, Austin Troy, and Lisa Aultman-Hall.We are grateful for funding from the MITRE Corporation, the UVM Transportation Research Center, the Vermont Complex Systems Center, and the Vermont Advanced Computing Core, which was supported by NASA (NNX 08A096G). PSD was supported by NSF CAREER Grant No. 0846668.

\noindent{\textbf{Author Contributions}} \\
\indent  C.M.D. and P.S.D. designed the research, M.R.F. prepared the figures, and C.M.D. and M.R.F. wrote the manuscript. M.R.F., L.M., P.S.D., and C.M.D analyzed the data and reviewed the manuscript.

\noindent{\textbf{Competing Financial Interests}}\\
\indent The authors declare no competing financial interests.

\clearpage

\noindent{\textbf{\textit{Supplementary Material}: \\ Happiness and the Patterns of Life: A Study of Geolocated Tweets} \\
\textbf{Morgan R. Frank, Lewis Mitchell, Peter S. Dodds, \\ Christopher M. Danforth\\
{\footnotesize  Computational Story Lab, Department of Mathematics and Statistics, Vermont Complex Systems Center, Vermont Advanced Computing Core,}\\
{\footnotesize University of Vermont, Burlington, Vermont, United States of America }}

\setcounter{figure}{0} \renewcommand{\thefigure}{S\arabic{figure}}
\setcounter{table}{0} \renewcommand{\thetable}{S\arabic{table}}

\begin{figure}[!htdp]
\begin{center}
	\includegraphics[scale=1]{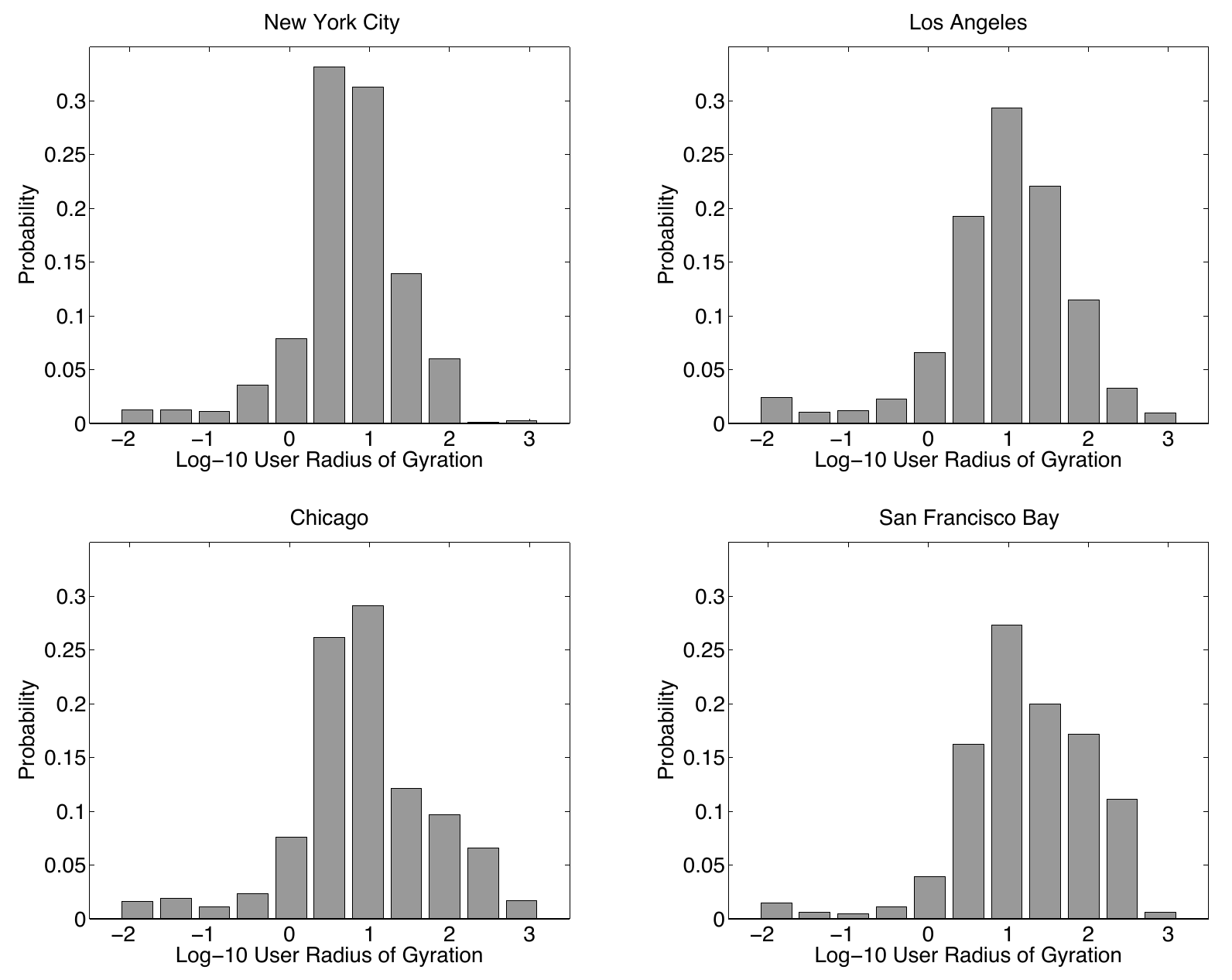}
\end{center}
\caption{The distributions of gyradius (km) for four cities appear to be approximately lognormal. The mode distance (binned) is larger for Los Angeles and San Francisco than for Chicago and New York City. We note that these distributions were calculated for all individuals whose expected location fell within the latitude and longitude bounds of main text Fig. 2, and thus reflect a modified set of individuals than those identified with cities in Fig. \ref{Figpopulation}.}
\label{Figraddist} 
\end{figure}

\begin{figure*}[!htdp]
\begin{center}
	\includegraphics[scale=1]{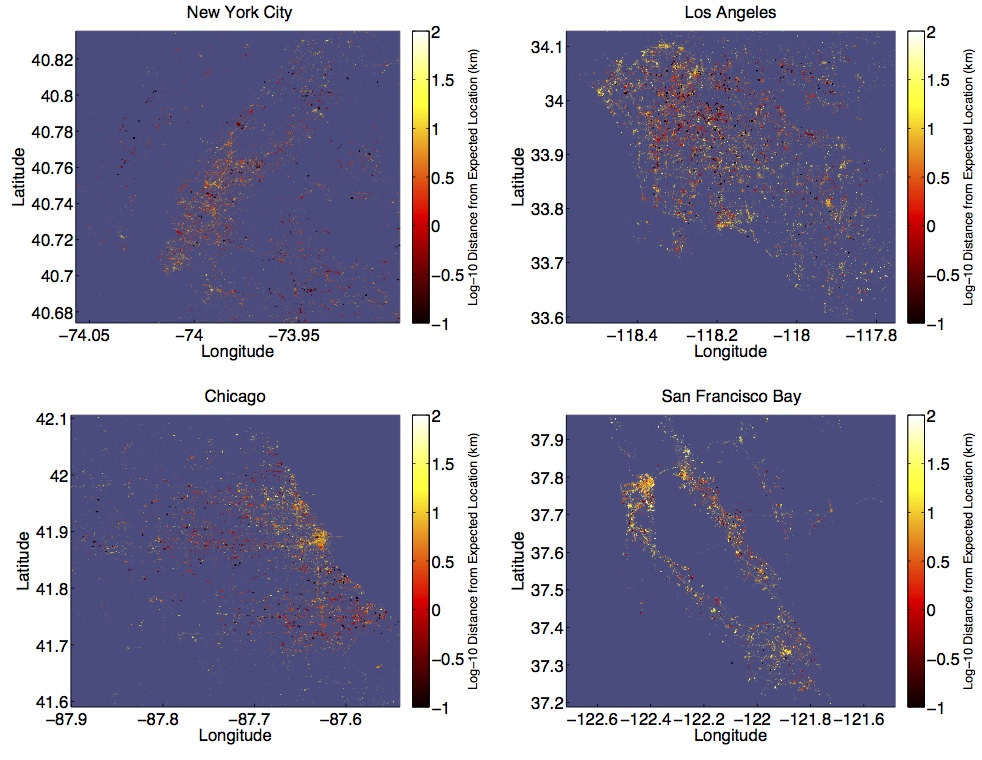}
\end{center}
\caption{The distance from expected location, calculated for each individual, is shown for each tweet authored in four example cities in 2011. Spatial clustering is observed (Table S2); messages authored by individuals far from their expected location are more likely to appear close to each other. The number of tweets shown for each city is $N=56,650$ (Chicago), $N=103,213$ (Los Angeles), $N=42,089$ (New York City), and $N=45,754$ (San Francisco).}
\label{Figcitydots2} 
\end{figure*}

\clearpage

\begin{table}[!htdp]
\begin{center}
\begin{tabular}{|c|c|c|c|}
\hline
City & Geary's C (p-value) & Moran's I (p-value) & Tweets \\ \hline
Chicago & 0.43 $(<10^{-15})$ & 0.27 $(<10^{-15})$ & 56650\\
Los Angeles & 0.47 $(<10^{-15})$ & 0.22 $(<10^{-15})$ & 103213\\
New York City & 0.40 $(<10^{-15})$ & 0.34 $(<10^{-15})$ & 42089 \\
San Francisco Bay & 0.37 $(<10^{-15})$ & 0.46 $(<10^{-15})$ & 45754 \\
\hline
\end{tabular}
\end{center}
\caption{Evidence for clustering is observed in both Geary's C and Moran's I spatial autocorrelation for tweet location colored by gyradius (Figure 2).}
\label{Taspatial1}
\end{table}%

\begin{table}[!htdp]
\begin{center}
\begin{tabular}{|c|c|c|c|}
\hline
City & Geary's C (p-value) & Moran's I (p-value) & Tweets  \\ \hline
Chicago & 0.74 $(<10^{-15})$ & 0.14 $(<10^{-15})$ & 56650 \\
Los Angeles & 0.64 $(<10^{-15})$ & 0.16 $(<10^{-15})$ & 103213\\
New York City & 0.65 $(1.3 \times 10^{-3})$ & 0.07 $(<10^{-15})$ & 42089 \\
San Francisco Bay & 0.55 $(<10^{-15})$ & 0.34 $(<10^{-15})$ & 45754 \\
\hline
\end{tabular}
\end{center}
\caption{Evidence for clustering is observed in Geary's C and Moran's I spatial autocorrelation for tweet distance from expected location as well (Figure S2).}
\label{Taspatial2}
\end{table}%

\begin{table}[!htdp]
\begin{center}
\begin{tabular}{|c|c|c|c|}
\hline
City & Geary's C (p-value) & Moran's I (p-value) & Individuals \\ \hline
Chicago & 0.43 $(<10^{-15})$ & 0.60 $(<10^{-15})$ & 563\\
Los Angeles & 0.29 $(<10^{-15})$ & 0.70 $(<10^{-15})$ & 983\\
New York City & 0.21 $(<10^{-15})$ & 0.75 $(<10^{-15})$ & 387 \\
San Francisco Bay & 0.52 $(2.3\times10^{-12})$ & 0.44 $(<10^{-15})$ & 423 \\
\hline
\end{tabular}
\end{center}
\caption{Evidence for clustering is observed in Geary's C (local) and Moran's I (global) spatial autocorrelation calculated for mode location colored by gyradius (not shown to preserve privacy). Note that Geary's C values fall between 0 and 2, with 1 indicating no correlation and values smaller than 1 suggesting increasing correlation. Moran's I values range from -1 to 1, with larger values suggesting positive correlation.}
\label{Taspatial3}
\end{table}%

\clearpage

\begin{figure}[!htdp]
\begin{center}
	\includegraphics[width=\textwidth]{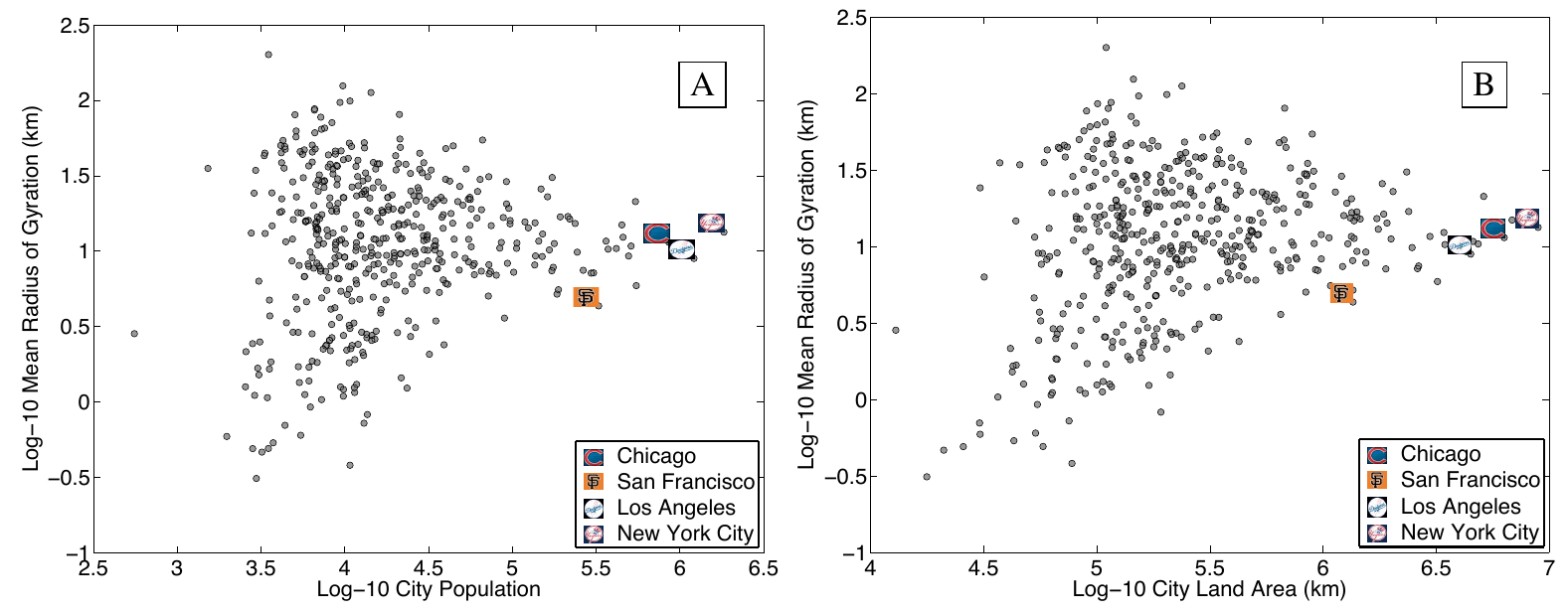}
\end{center}
\caption{The mean gyradius of individuals whose expected location falls within each city is plotted against the city's population (A) and land area (B). Shown are cities containing at least 50 individuals with a nonzero gyradius, each individual having authored at least 30 geolocated tweets. City boundaries are defined by \cite{census} which encompasses a smaller area for the four cities illustrated in the main text Fig. 2. Generally, gyradius increases with city population and land area, with no large cities exhibiting a small mean radius. Pearson correlations: Population $\rho=0.10,\ p=0.03$, Land Area $\rho=0.24, \ p=2\times10^{-7}$.}
\label{Figpopulation}
\end{figure}

\begin{table}[!htdp]
\begin{center}
\begin{tabular}{|c|c|c|}
\hline
rank & radius (km) & city \\ \hline
1 & 200.6 & Martinsville, VA \\
2 & 124.5 & Middletown, OH \\
3 & 112.3 & Elkhart, IN \\
4 & 98.8 & Pottstown, PA \\
5 & 96.6 & Decatur, IL \\
$\cdots$ & $\cdots$ & $\cdots$ \\
215 & 13.3 & New York City, NY \\
247 & 11.4 & Chicago, IL \\
300 & 8.94 & Los Angeles, CA \\
387 & 4.33 & San Francisco \& Oakland, CA \\
$\cdots$ & $\cdots$ & $\cdots$ \\
468 & 0.492 &  Greenville, MS \\
469 & 0.491 & Athens, OH \\
470 & 0.465 &  Key West, FL \\
471 & 0.381 &  El Centro Calexico, CA \\
472 & 0.312 & Pullman, WA \\ \hline
\hline
\end{tabular}
\end{center}
\caption{Top and Bottom 5 cities with respect to mean gyradius, along with the four cities investigated in main text Fig 2.}
\label{Tacities}
\end{table}%


\clearpage

\begin{figure}[!htdp]
\begin{center}
	\includegraphics[width=.5\textwidth]{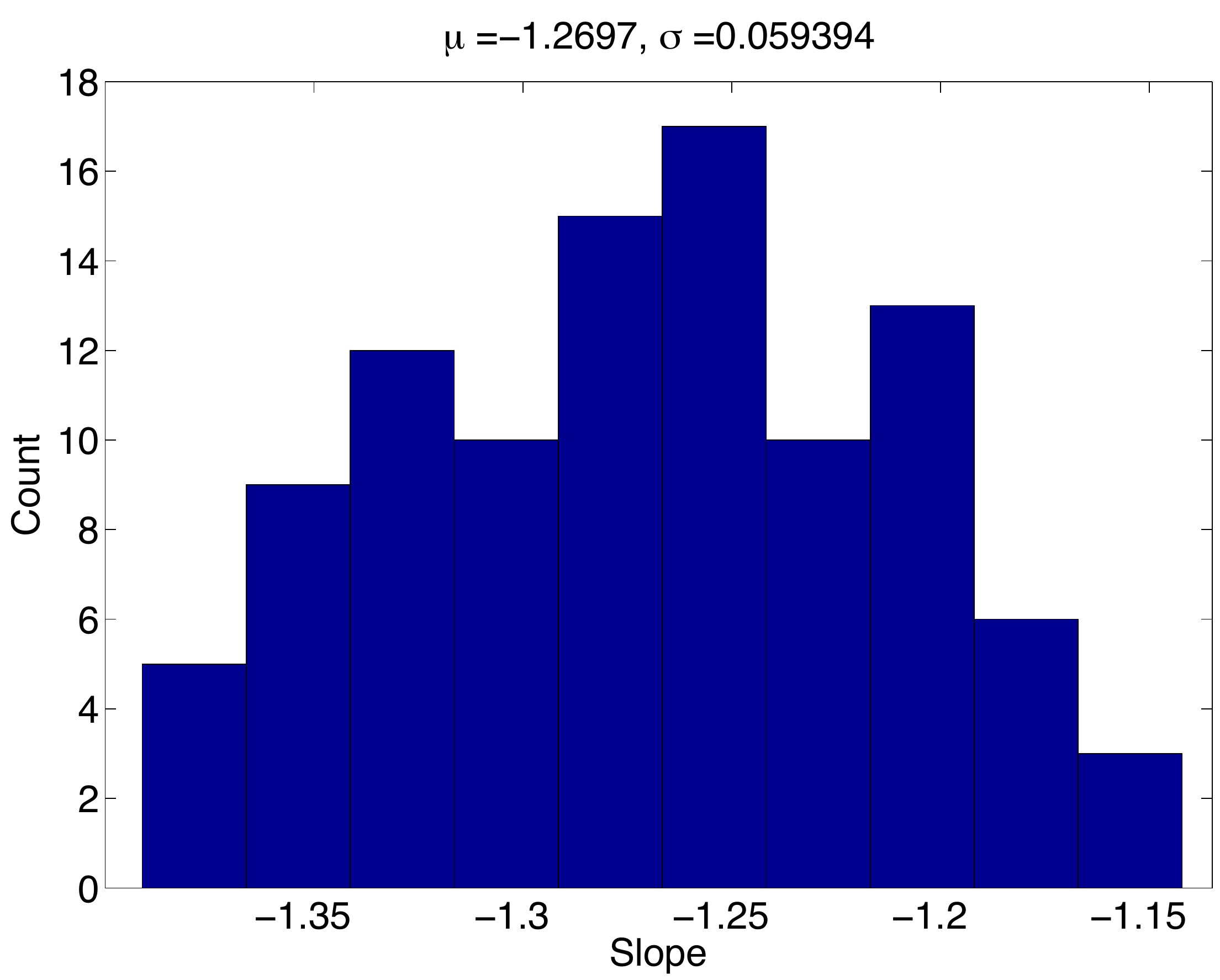}
\end{center}
\caption{A random 10\% of individuals (30 out of 300) are removed from Figure 5A, and the slope of the probability fit (red curve) is recalculated. Repeating the procedure 100 times, we find the above distribution of slopes. The mean of this distribution agrees well with that reported in Figure 5A. Additionally, fitting the power law model to the leading 10 locales, using only individuals who have at least 10 locales, we also get a slope of roughly $-1.3$ (not shown).}
\label{Figzipfstats}
\end{figure}

\clearpage

\begin{figure*}[!htdp]
\begin{center}
	\includegraphics[scale=1]{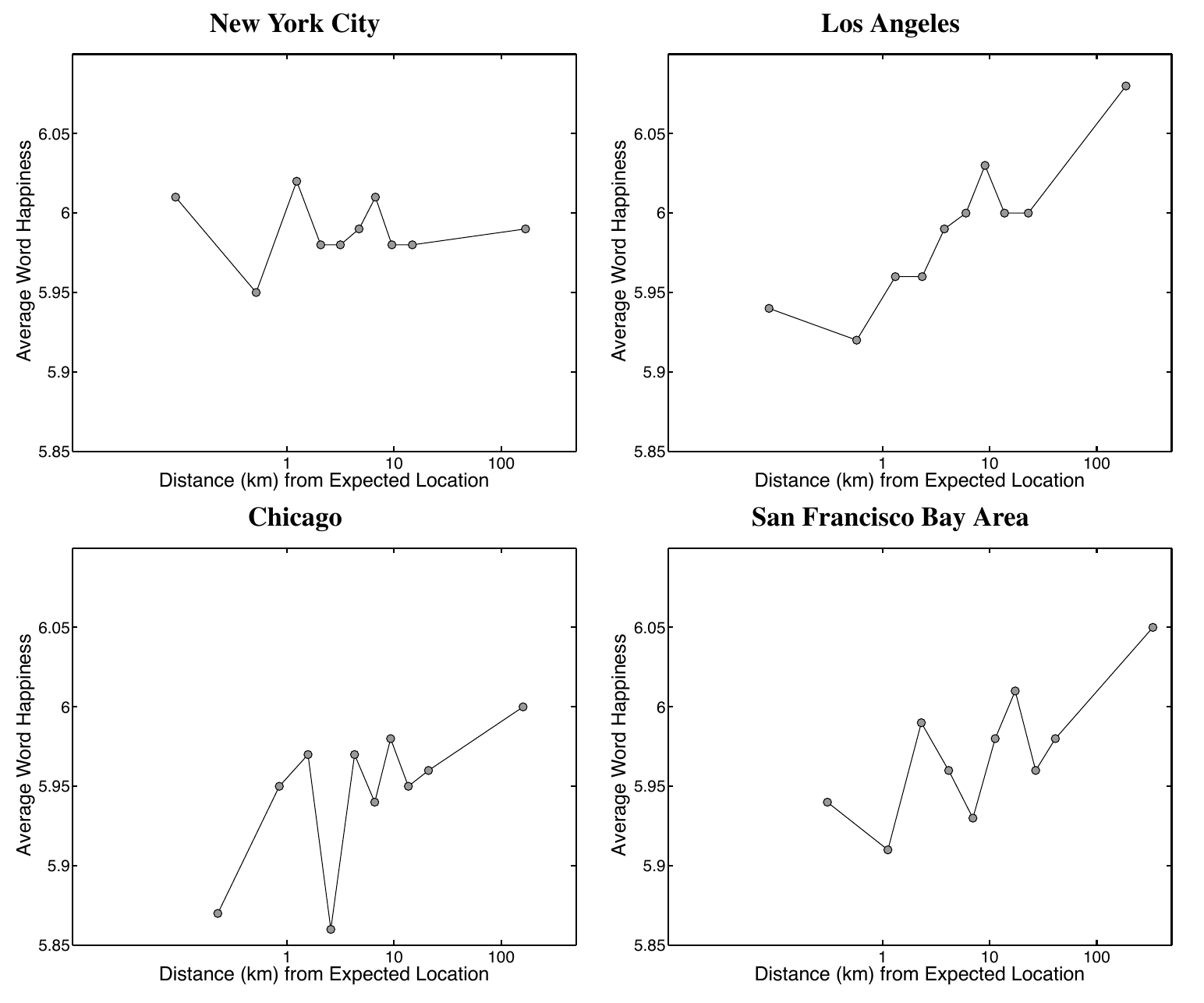}
\end{center}
\caption{For New York City, Los Angeles, Chicago, and the San Francisco Bay Area, we group messages into equally sized bins by the distance from expected location of their author, and measure the average word happiness of each group. These plots exhibit similar trends to that observed in main text Fig. 6A with the exception of New York City.}
\label{Fighappy1} 
\end{figure*}

\begin{figure*}[!htdp]
\begin{center}
	\includegraphics[scale=1]{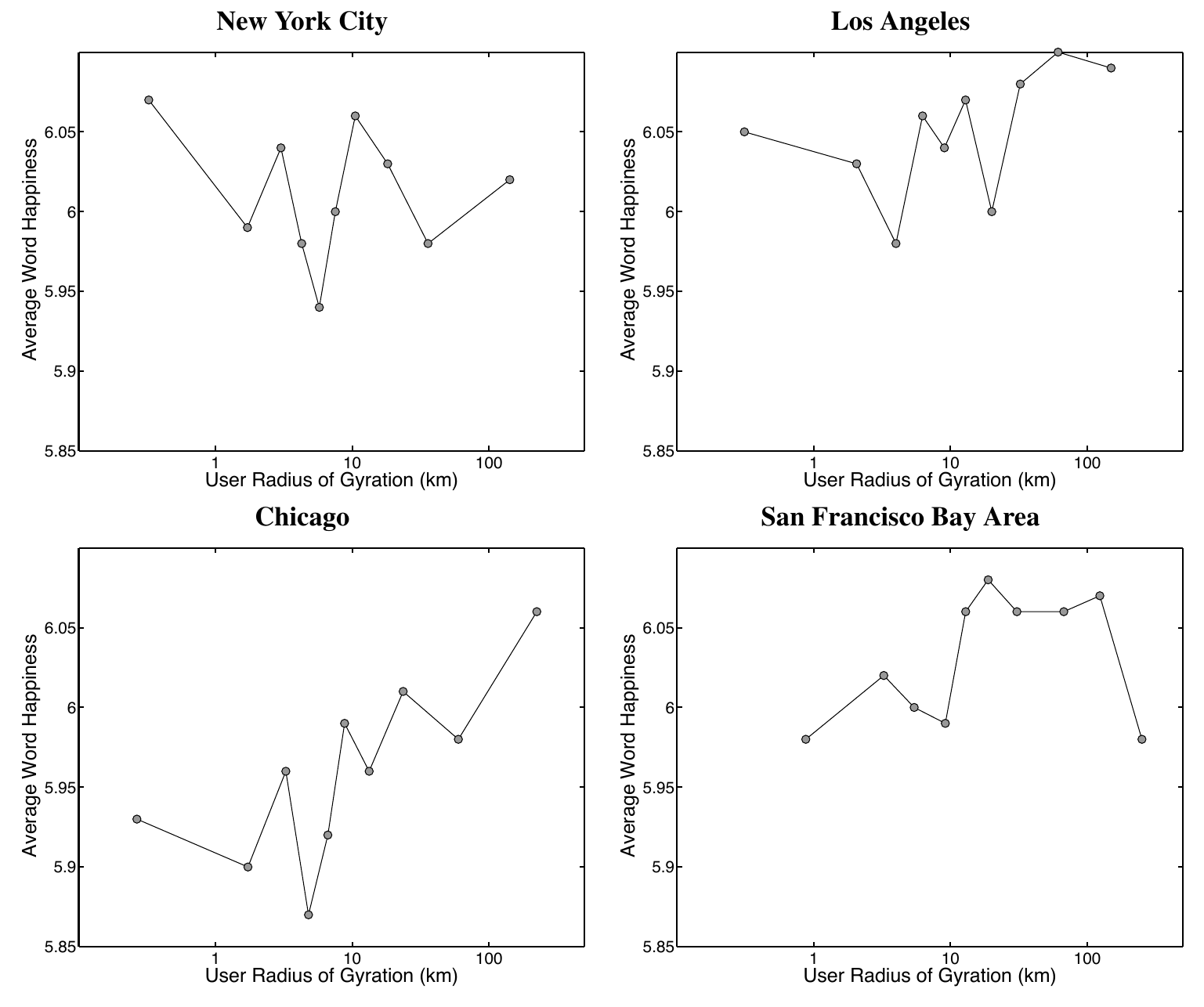}
\end{center}
\caption{For New York City (130 individuals/bin), Los Angeles (175 individuals/bin), Chicago (125 individuals/bin), and the San Francisco Bay Area (63 individuals/bin), we group individuals into equally sized bins by their gyradius and measure the average word happiness of each group. These plots exhibit similar trends to that observed in main text Fig. 6B with the exception of the largest radius group in the San Francisco Bay Area, and New York City as a whole.}
\label{Fighappy2} 
\end{figure*}

\begin{figure*}[!htdp]
\begin{center}
\begin{multicols}{2}
	\begin{overpic}[width=.9\columnwidth]{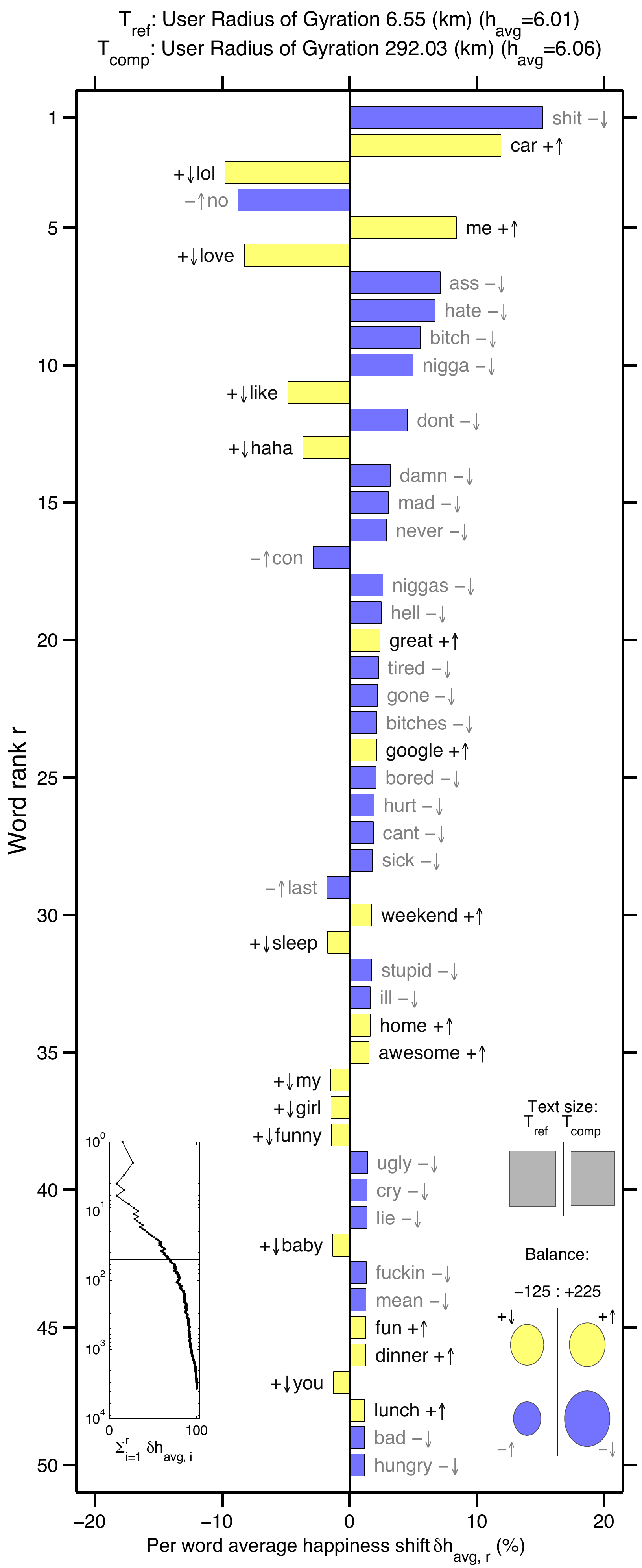}
		\put(10,100){\fbox{A}}
	\end{overpic}
	\begin{overpic}[width=.9\columnwidth]{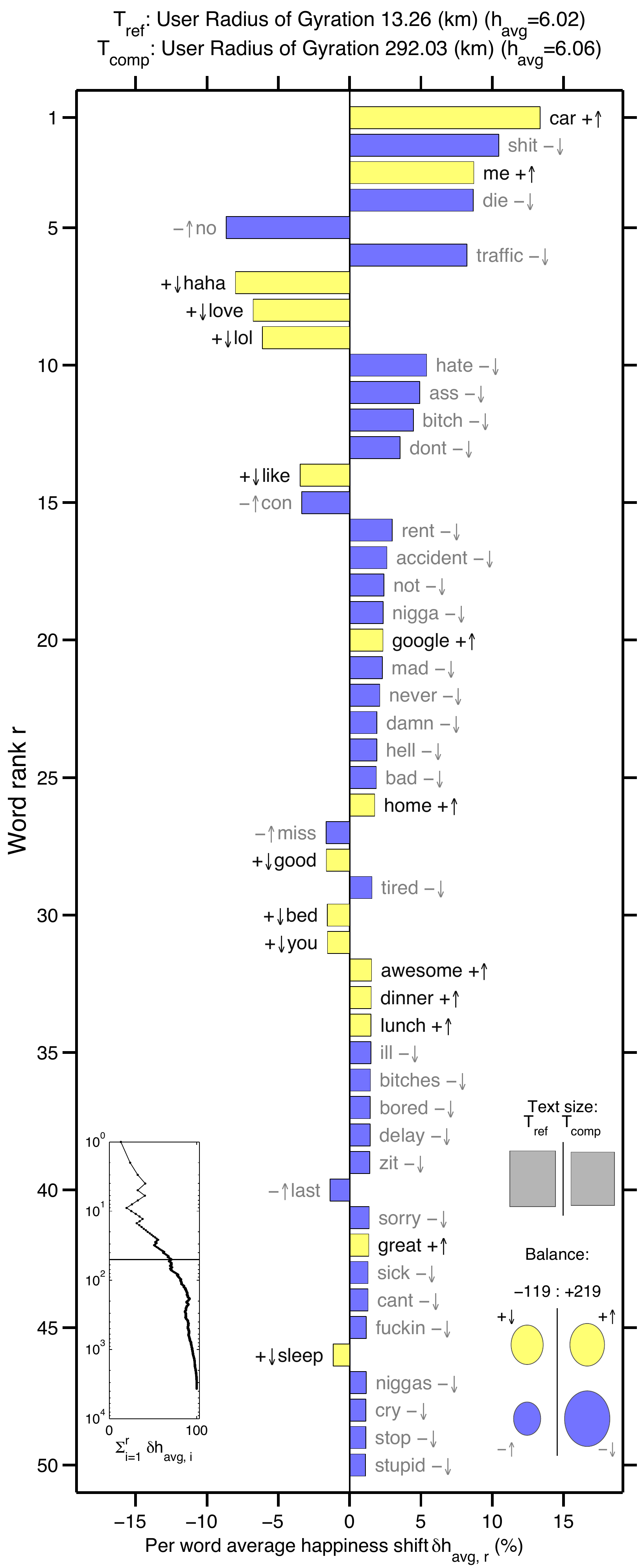}
		\put(10,100){\fbox{B}}
	\end{overpic}
\end{multicols}
\end{center}
\caption{We compare the 6.55 km gyradius group versus the 292.03 km gyradius group (A). We find that the 292.03 km group has relatively frequent use of the words `car' and `weekend' suggesting that this group travels on the weekends perhaps to a vacation home as suggested by use of the word `home'. (B) We compare the 13.26 km gyradius group versus the 292.03 km gyradius group. We find that the 292.03 km group uses the word `car' more frequently than the 13.26 km group which, interestingly, uses the word `traffic' more frequently. Again the increased relative usage of these words seems fitting for a groups with these patterns of movement.}
\label{Figwords1} 
\end{figure*}

\begin{figure*}[!htdp]
\begin{center}
\begin{multicols}{2}
	\begin{overpic}[width=.9\columnwidth]{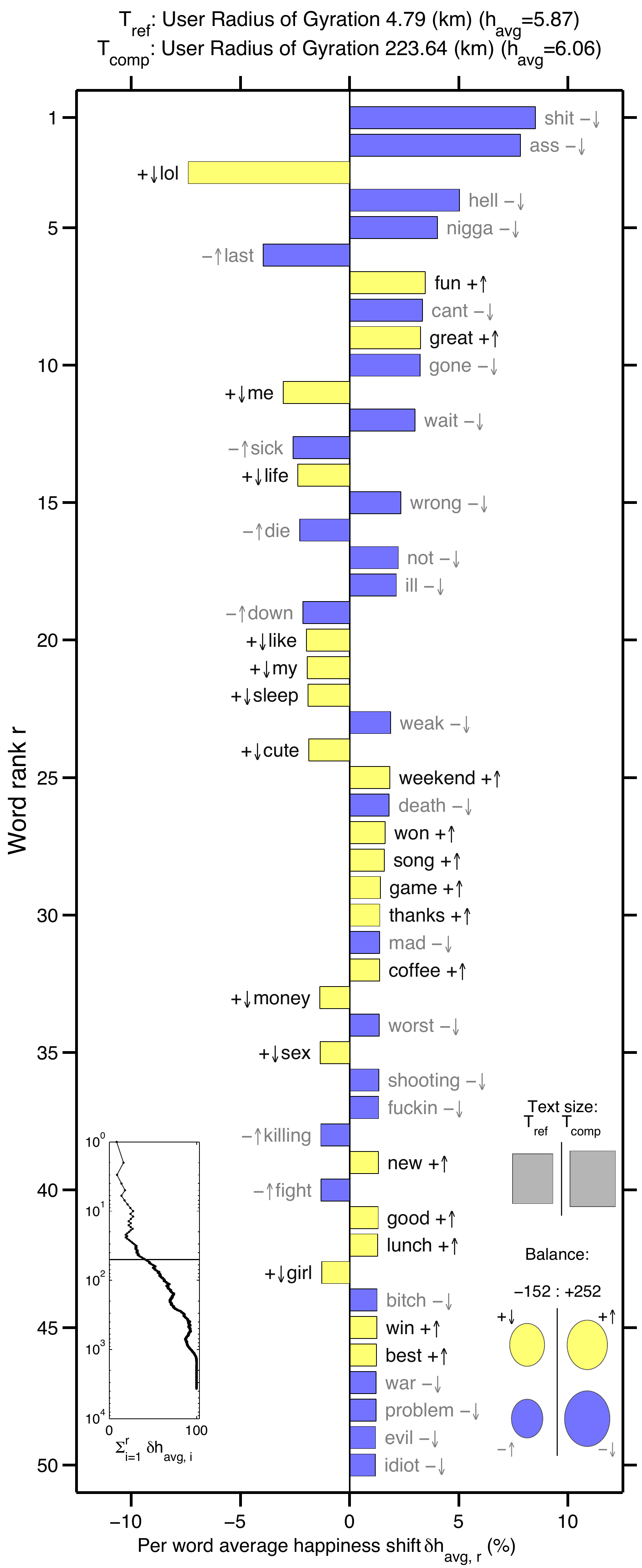}
		\put(10,100){\fbox{A}}
	\end{overpic}
	\begin{overpic}[width=.9\columnwidth]{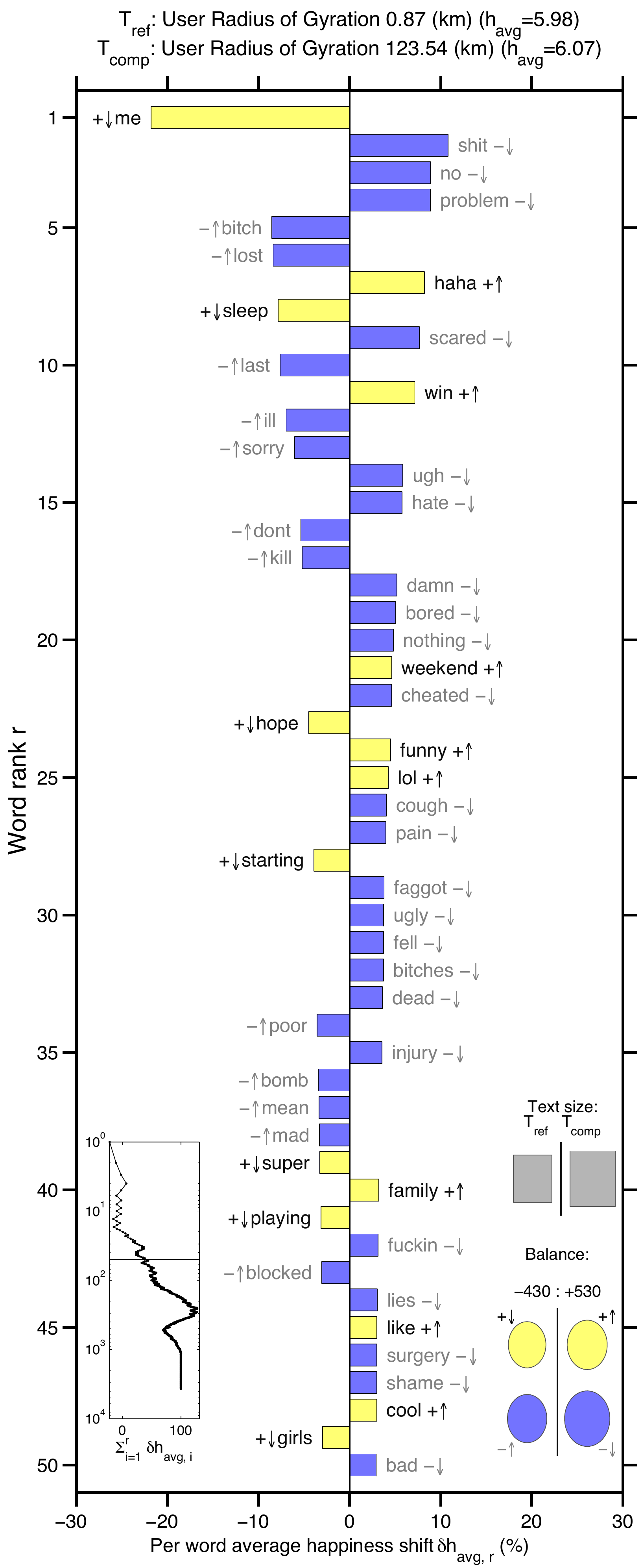}
		\put(10,100){\fbox{B}}
	\end{overpic}
\end{multicols}
\end{center}
\caption{(A) A word shift comparing the 4.79 km gyradius group to the 223 km gyradius for Chicago. We observe the first group is less happy because of increased usage of profanity and negative words like `can't', `gone', and 'wrong'. (B) A word shift comparing the .87 km gyradius group to the 123.54 km gyradius group for the San Francisco Bay Area. We find the second group to be happier because of an increase in positive words like `haha', `win', `weekend', `funny', and `lol', along with a decrease in negative words like `no', `problem', and `hate'.}
\label{Figwords2} 
\end{figure*}

\begin{figure*}[!htdp]
\begin{center}
	\includegraphics[scale=.45,trim=1cm 6cm 0cm 6cm,clip]{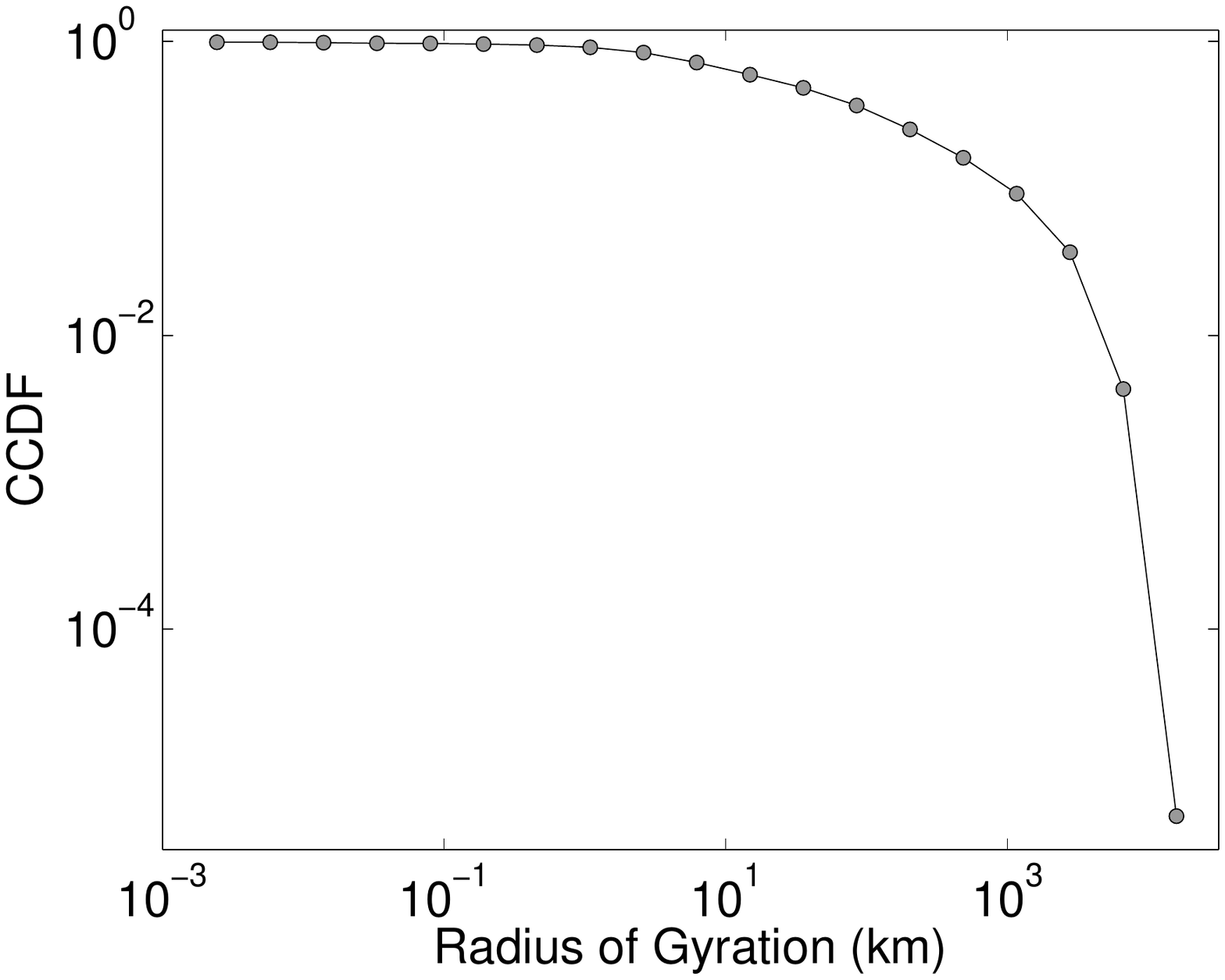}
\end{center}
\caption{Complementary Cumulative Distribution Function (CCDF) for the gyradii of all users with at least 30 geolocated messages. Gonzalez \cite{movement} found this distribution to be well modeled by a truncated power law with an exponential tail.}
\label{Figuser5b} 
\end{figure*}

\clearpage

\textbf{Normalizing Human Trajectory}

\indent To compare the shape of trajectories of individuals traveling in different directions and over different distances, we use the methods introduced by Gonz\'{a}lez et al. \cite{movement}. We will examine the normalization steps for two individuals we will call user A and user B. We have 768 geolocated tweets for user A and 1,882 geolocated tweets for user B. User A has gyradius $r^{A}=463.61$ km and user B has gyradius $r^{B}=54.28$ km. Fig. \ref{Figuser1} represents the geospatial tweet locations for user A and user B, but we have shifted their coordinate system to maintain their anonymity. We have also allowed for a slight spatial separation between the locations for user A and the locations of user B for clarity.\\

\begin{figure}[!htdp]
\begin{center}
	\includegraphics[width=.45\textwidth]{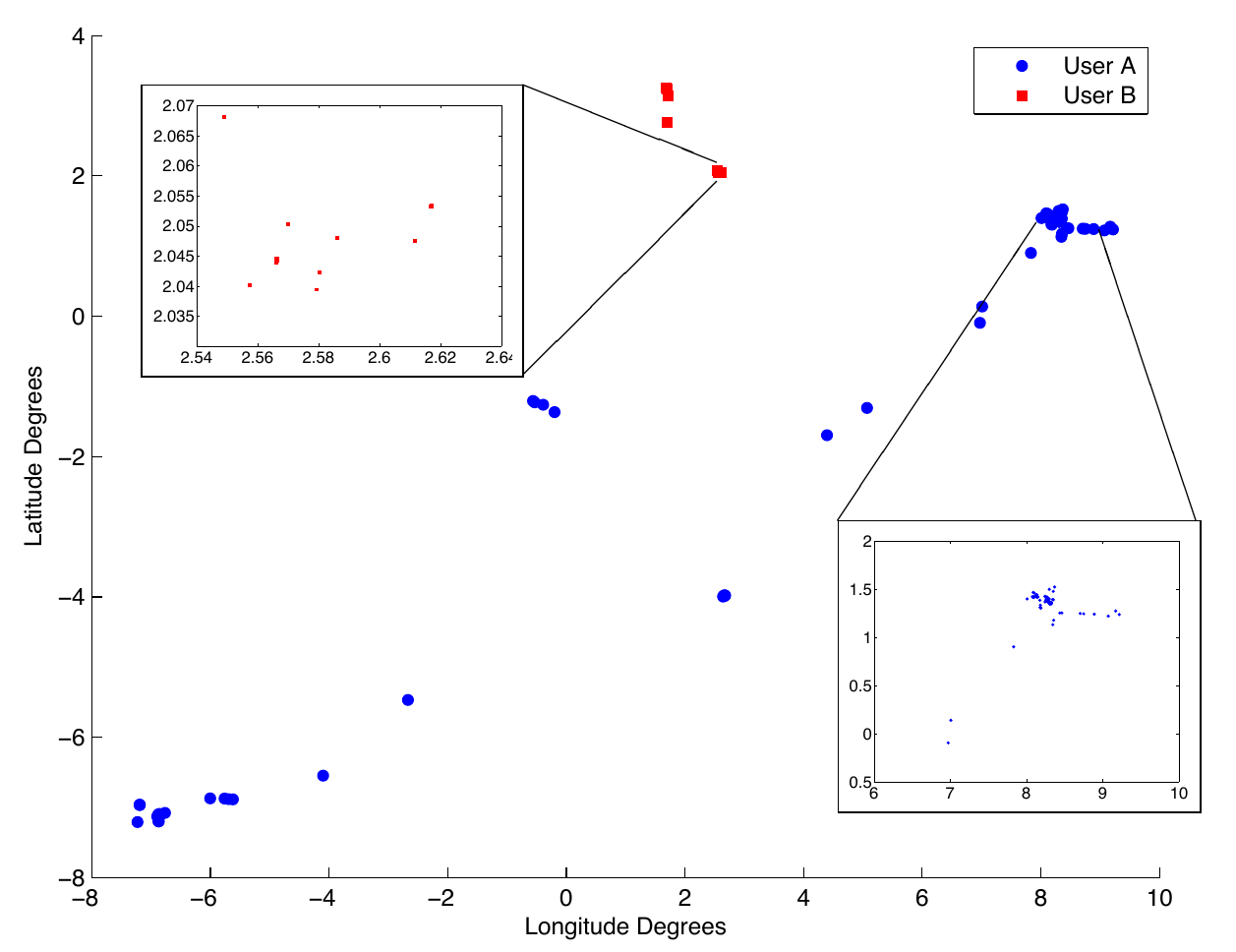}
\end{center} 
\caption{Tweet locations for User A and User B.}
\label{Figuser1} 
\end{figure}

\indent In Fig. \ref{Figuser2}, we apply the linear transformation shifting each location for the user to the distance in kilometers from their center of mass, i.e. the expected location of the user. The difference in gyradius between user A and user B is still very apparent in the axis ranges for this plot. Notice that the directional relationships between the tweet locations for each user have still been preserved. We can see that user A travels predominantly in a southwest direction, while user B travels primarily in a northwest direction.

\begin{figure}[!htdp]
\begin{center}
	\includegraphics[scale=.45,trim=2cm 9cm 0cm 7cm,clip]{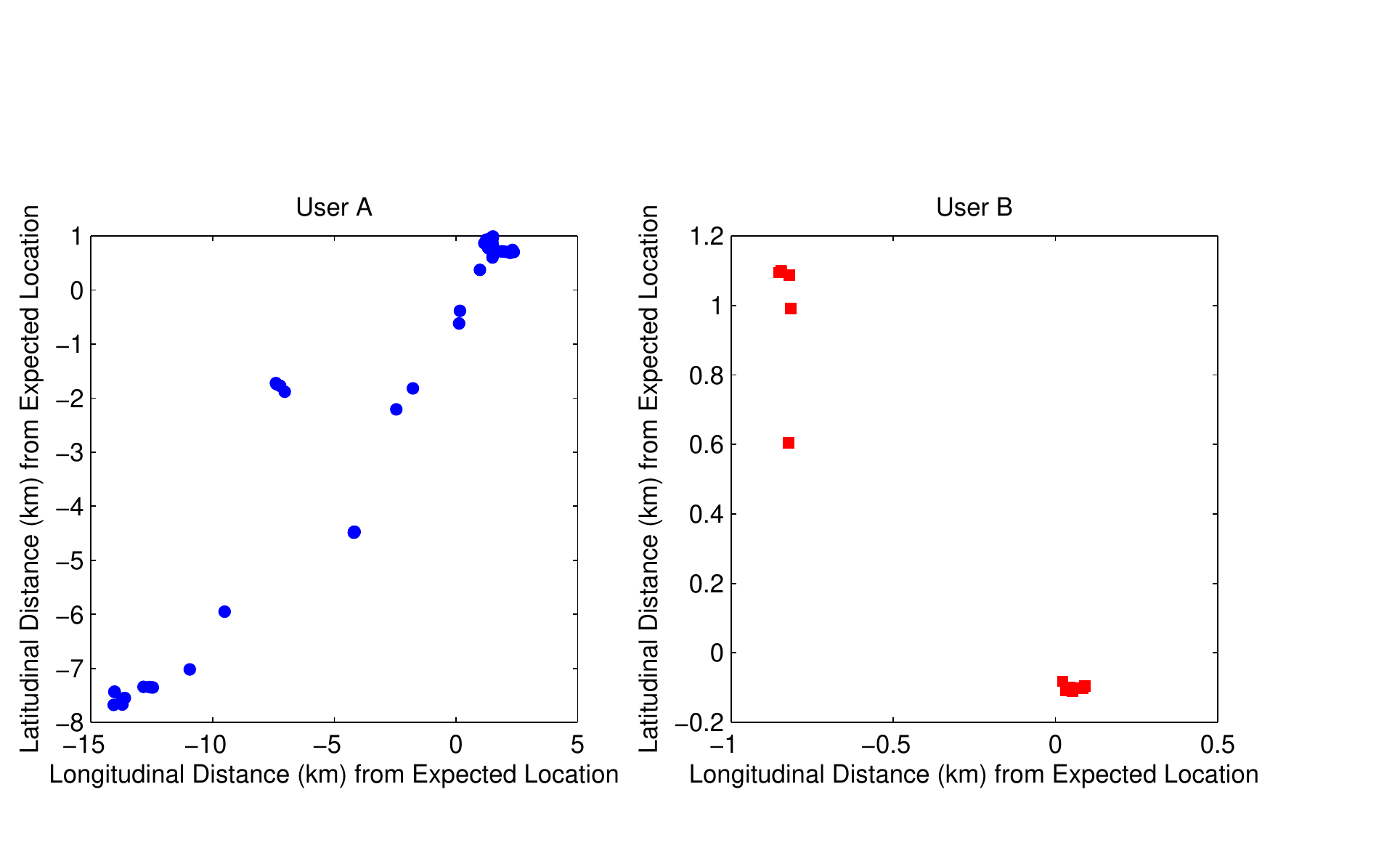}
\end{center}
\caption{Tweet locations for User A and User B transformed to the distance in kilometers from their expected locations, respectively.}
\label{Figuser2} 
\end{figure}

\indent To normalize for direction of travel, let the set of tweet locations for user $i$ be represented by the set of equally weighted masses at each of the tweet locations $\{(x_{1},y_{1}),(x_{2},y_{2}),\dots,(x_{n},y_{n})\}$. Now we calculate the tensor of inertia ($I$) for each set of weighted $(x,y)$-points as
\begin{equation*}
	I=\begin{bmatrix}
		\displaystyle\sum_{j=1}^{n}x_{j}^{2} & -\displaystyle\sum_{j=1}^{n}x_{j}y_{j} \\
		 -\displaystyle\sum_{j=1}^{n}x_{j}y_{j} & \displaystyle\sum_{j=1}^{n}y_{j}^{2}
		\end{bmatrix}
\end{equation*}
The eigenvector of $I$ corresponding to the largest eigenvalue of $I$ represents the direction along which most of user $i$'s trajectory occurs; we call this the principal axis for user $i$ (see Fig. \ref{Figuser3}).

\begin{figure}[!htdp]
\begin{center}
	\includegraphics[scale=.45,trim=2cm 7cm 0cm 7cm,clip]{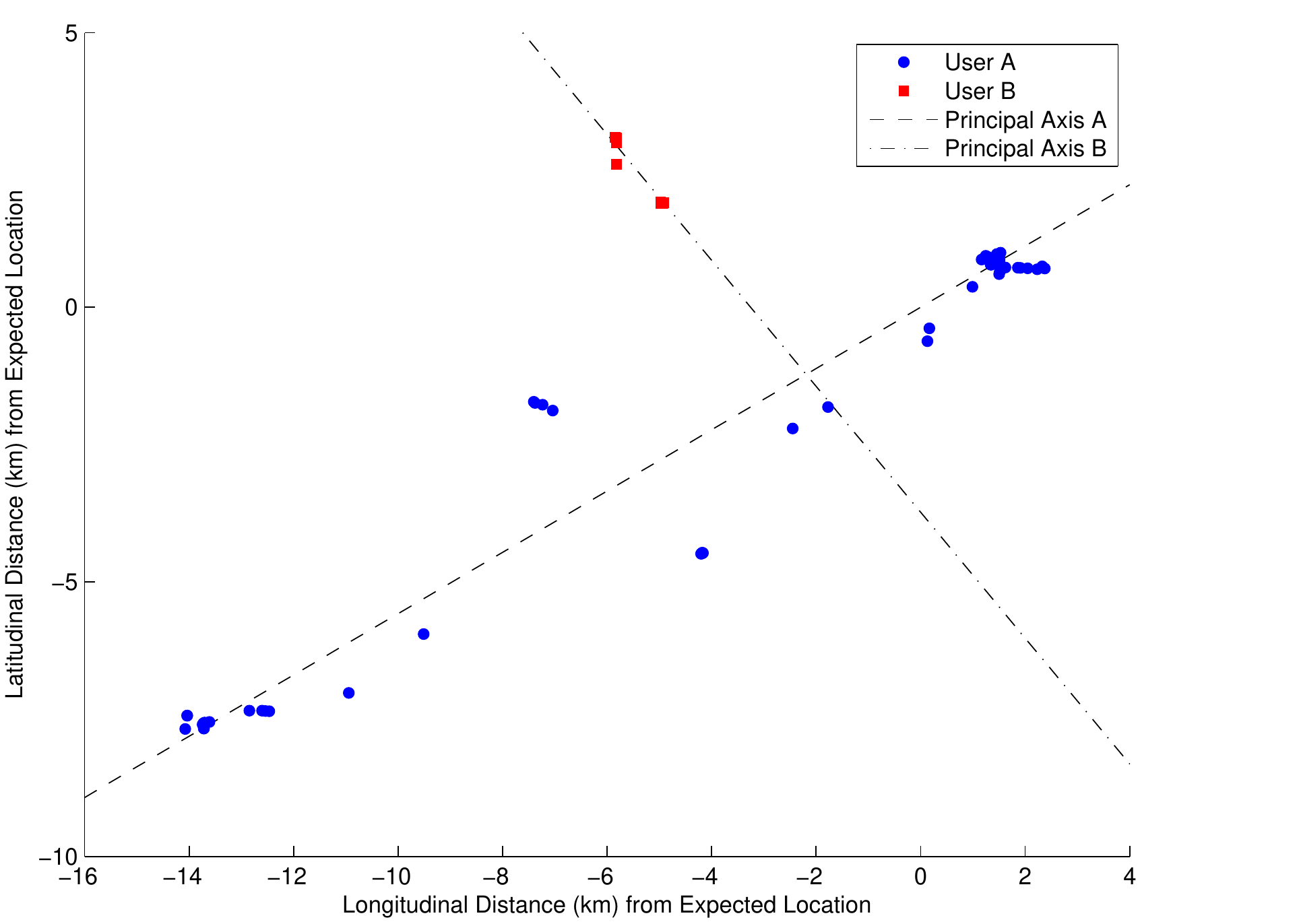}
\end{center}
\caption{The tweet locations for User A and User B along with a line representing the principal axis for that user.}
\label{Figuser3} 
\end{figure} 

Now we can determine the angle necessary to rotate the set of points for user $i$ so that the the resulting principal axis is the $x$-axis. Fig. \ref{Figuser4} shows the results of this step. We see that the principal axis for user A and user B is now the $x$-axis.

\begin{figure}[!htdp]
\begin{center}
	\includegraphics[scale=.45,trim=2cm 7cm 0cm 7cm,clip]{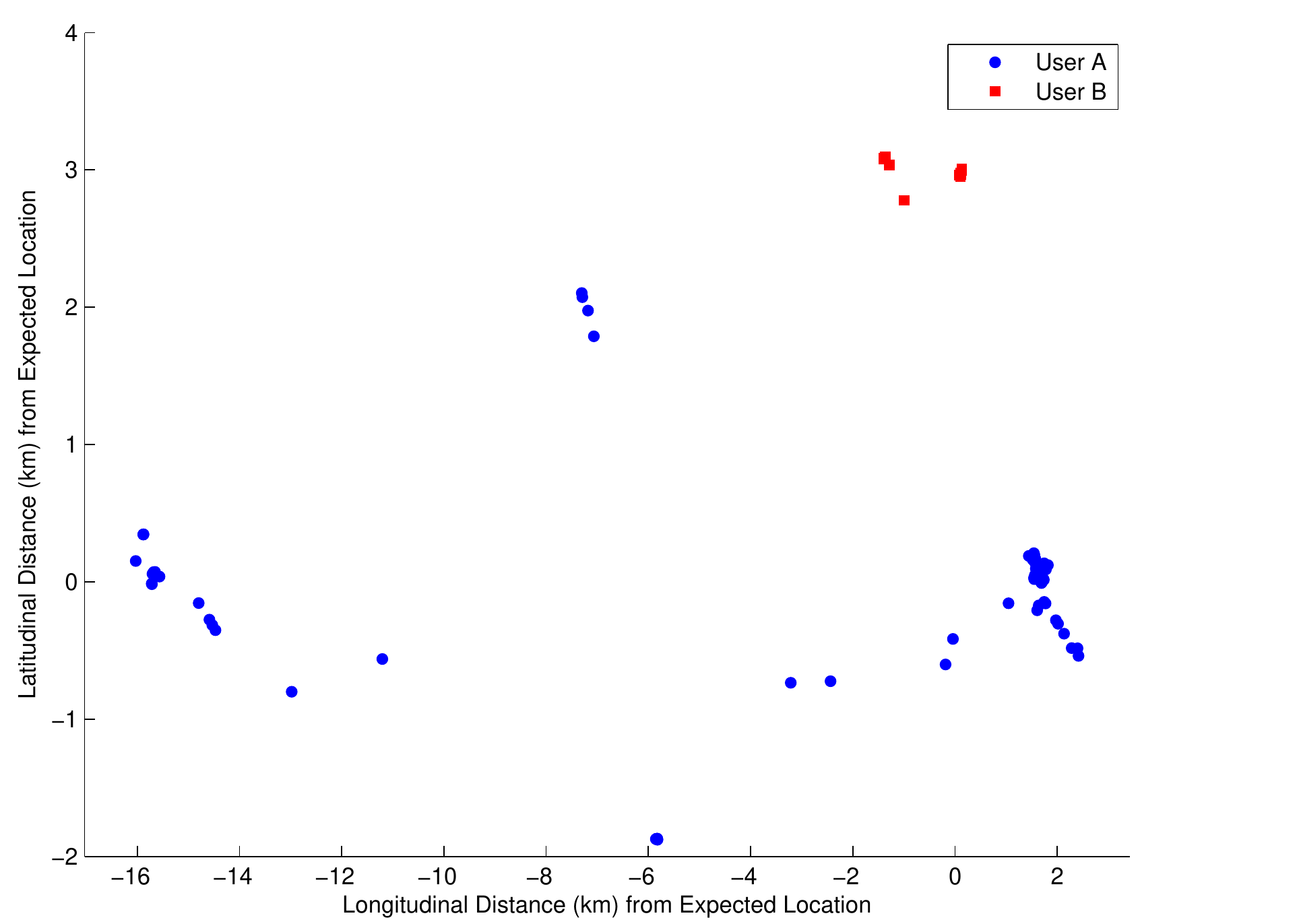}
	\end{center}
\caption{The results after rotating the locations of User A and User B. We see that they now both have principal axes of trajectory pointing due west.}
\label{Figuser4} 
\end{figure}

\indent The final step is to normalize for individuals with different gyradius. We accomplish this by dividing the $x$-coordinate of each rotated tweet location for user $i$ by $\sigma_{x}$, where $\sigma_{x}$ is the standard deviation of the $x$-coordinates of the rotated tweet locations for user $i$, and similarly dividing by $\sigma_{y}$ for the $y$-coordinates. The final result is shown in Fig. \ref{Figuser5}.

\begin{figure}[!htdp]
\begin{center}
	\includegraphics[scale=.45,trim=2cm 7cm 0cm 7cm,clip]{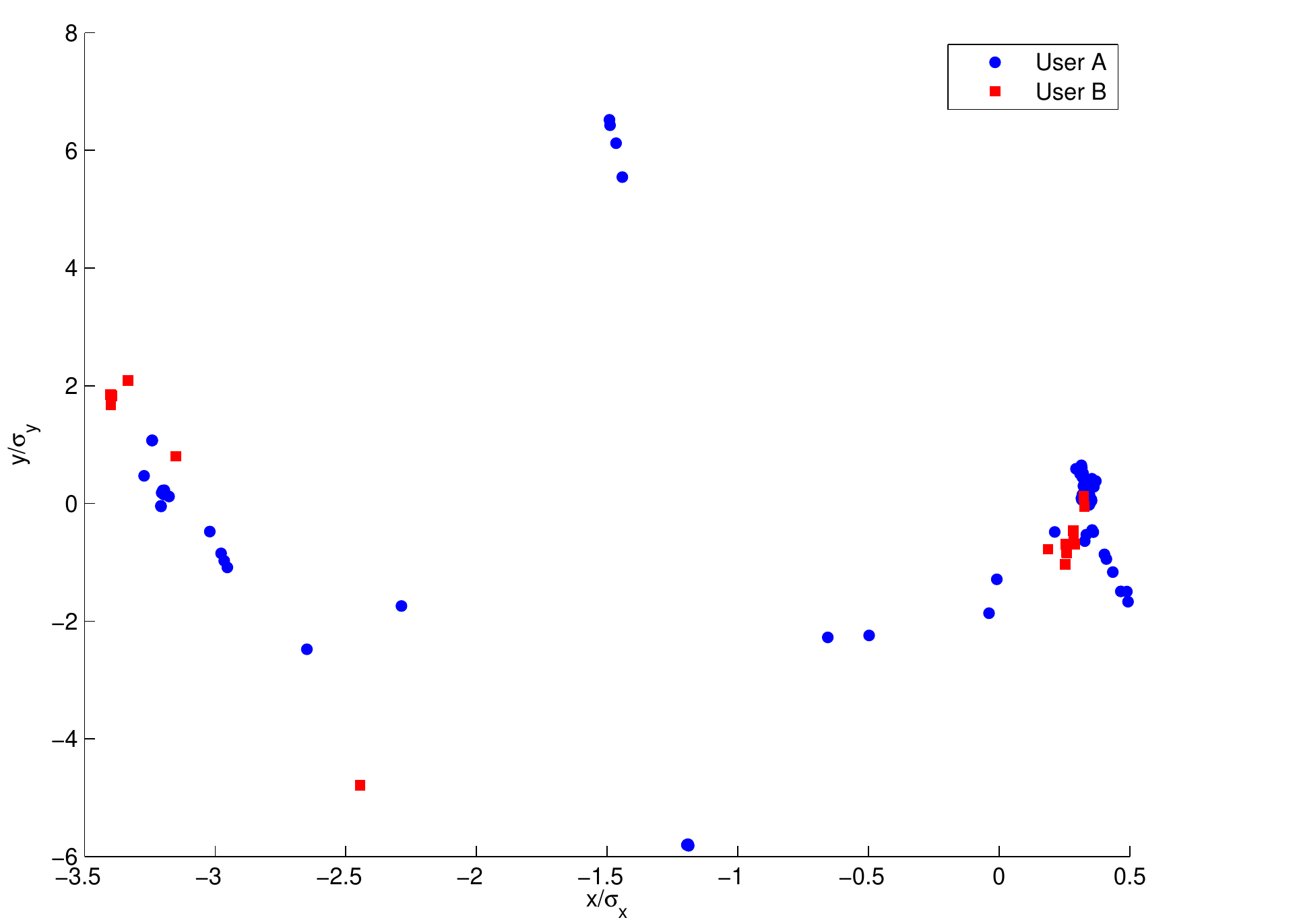}
\end{center}
\caption{The rotated tweet locations of User A and User B after normalizing for gyradius. The origin represents the center of mass of the respective individuals' trajectory, namely $\langle \vec{p}^{a}\rangle $ from equation (2).}
\label{Figuser5} 
\end{figure}

\indent As a result, we can compare the shape of the trajectories for User A and User B having normalized for direction and gyradius. We can see that both User A and User B have most of their  normalized tweet locations in two main clusters.\\

\end{document}